\documentclass{aastex631}
\usepackage{bm}

\usepackage{textcomp}
\newcounter{num}
\newcommand{\Rnum}[1]{\setcounter{num}{#1} \Roman{num}}

\shorttitle{Evolution of the Angular Momentum of Molecular Cloud Cores}
\shortauthors{Misugi et al.}

\graphicspath{{./}{figures/}}

\begin{document}

\title{Evolution of the Angular Momentum of Molecular Cloud Cores in Magnetized Molecular Filaments}

\author{Yoshiaki Misugi}
\affiliation{Graduate Schools of Science and Engineering, Kagoshima University, Kagoshima, Japan}
\affiliation{Division of Science, National Astronomical Observatory of Japan, Tokyo, Japan}

\author{Shu-ichiro Inutsuka}
\affiliation{Department of Physics, Graduate School of Science, Nagoya University, Nagoya, Japan}

\author{Doris Arzoumanian}
\affiliation{Division of Science, National Astronomical Observatory of Japan, Tokyo, Japan}

\author{Yusuke Tsukamoto}
\affiliation{Graduate Schools of Science and Engineering, Kagoshima University, Kagoshima, Japan}



\begin{abstract}
The angular momentum of molecular cloud cores plays a key role in the star formation process. However, the evolution of the angular momentum of molecular cloud cores formed in magnetized molecular filaments is still unclear. In this paper, we perform three-dimensional magnetohydrodynamics simulations to reveal the effect of the magnetic field on the evolution of the angular momentum of molecular cloud cores formed through filament fragmentation. As a result, we find that the angular momentum decreases by 30\% and 50\% at the mass scale of $1 \ {\rm M}_{\odot}$ in the case of weak and strong magnetic field, respectively. By analyzing the torques exerted on fluid elements, we identify the magnetic tension as the dominant process for angular momentum transfer for mass scales $\lesssim 3 \ {\rm M}_{\odot}$ for the strong magnetic field case. This critical mass scale can be understood semi-analytically as the timescale of magnetic braking. We show that the anisotropy of the angular momentum transfer due to the presence of strong magnetic field changes the resultant angular momentum of the core only by a factor of two. We also find that the distribution of the angle between the rotation axis and the magnetic field does not show strong alignment even just before the first core formation. Our results also indicate that the variety of the angular momentum of the cores are inherited from the difference of the phase of the initial turbulent velocity field. The variety could contribute to the diversity in size and other properties of protoplanetary disks recently reported by observations. 
\end{abstract}



\section{Introduction}\label{sec:intro}
The angular momentum of molecular cloud cores plays an essential role in the star formation process, since it is an important parameter to determine whether a multiple system is formed or not resulting from the fragmentation of the molecular cloud core \citep{Machida2008}. In the context of the evolution of the protobinary system, the condition of the formation of the circumbinary disk depends on the amplitude of the angular momentum of infalling gases \citep{Bate1997}. Recent studies showed that the angular momentum distribution of the core influences the accretion rate onto the protobinary system since the gas with small angular momentum falls onto the protobinary system along the polar direction, while the gas with high angular momentum falls onto the circumbinary disk if the condition of the formation of circumbinary disk is satisfied \citep{Matsumoto2019}. In addition, the structure of the angular momentum in a core is closely related to the evolution of the circumstellar disk because the infall gas brings the angular momentum onto the disk. Therefore, the angular momentum of a core and of its internal structure are important to understand the star and planet formation.\par
The magnetic field also plays an important role during the star formation process. The initial rotation and subsequent spin up due to the gravitational collapse of the core both have been shown to drive the outflow and jet by twisting the magnetic field lines \cite[e.g.,][]{Tomisaka2002,Machida2004,Machida2006, Banerjee2006, Hennebelle2008, Machida2008jet, Price2012, Tomida2013,Bate2014,Wurster2018}. The magnetic field transports the angular momentum due to the magnetic braking and makes a significant impact on the formation and evolution of circumstellar disks \cite[e.g.,][]{Machida2007,Machida2014,Price2007,Mellon2008,Dapp2010,Li2011,Dapp2012,Tsukamoto2015a,Tsukamoto2015b,Tomida2015,Tsukamoto2017}. Previous works also show that the angle between the rotation axis and the magnetic field direction affect the transfer of the angular momentum \citep{Mouschovias1985,Hennebelle2009,Joos2012,Li2013,Masson2016,Tsukamoto2018,Hirano2020}. \par
Herschel observations revealed that molecular filaments are the birthplaces of stars \citep{Andre2010,Arzoumanian_2011,Andre2014,Konyves2015,Pineda2022} and the filaments have an universal width of $\sim 0.1$ pc \citep{Arzoumanian_2011,Arzoumanian2019,Koch2015,Andre2022}. Moreover, observations show that prestellar cores and protostars are formed in thermally critical and supercritical filaments ($M_{\rm line} \gtrsim {\rm M_{\rm line,crit}}$) \citep{Andre2010,Tafalla_2015,Pineda2022}. $ {\rm M_{\rm line,crit}}=18 \ {\rm M_{\odot} pc^{-1}}$ for $T=10$ K denotes the critical line mass for isothermal unmagnetized cylindrical filaments \citep{Stodolkiewicz1963,Ostriker1964}. If the line mass is larger than the critical line mass, the filament cannot be in equilibrium against radial collapses without support from magnetic field and internal turbulence. Theoretical studies show that star forming cores are formed along a thermally critical/supercritical filament through self-gravitational fragmentation \citep{Inutsuka_Miyama1997}. Observational results from molecular line emissions revealed that the phase of centroid velocity fluctuations is shifted by $\lambda/4$ compared with that of the column density fluctuations along the filaments, where $\lambda$ represents a wavelength of the oscillation pattern \citep{Hacar2011, Shimajiri2023}. Since a convergence flow is needed to form the cores, the fact that the $\lambda/4$ shift is observed along the filaments supports the scenario in which cores are formed from filament fragmentation. In addition, recent NIR starlight polarization observations \cite[e.g.,][]{Kusune2019,Sugitani2019} and sub-mm dust continuum polarization observations \cite[e.g.,][]{Pillai2015,PlanckCollaboration2016,Cox2016,Doi2020,Arzoumanian2021} have confirmed that the filaments are magnetized with the magnetic field lines perpendicular to the longitudinal axes of  filaments. Therefore, studying the evolution of the cores formed from the gravitational fragmentation of magnetized filaments is important to understand the star formation process. \par
The angular momentum of cores formed in magnetized molecular clouds has been investigated using numerical simulations \citep{Dib2010,Kuffmeier2017,Chen2018, Ntormousi2019,Kuznetsova2020}. \cite{Dib2010} shows that the angular momentum of the cores measured in their simulations is smaller than the observed angular momentum. On the other hand, the core angular momentum measured in \cite{Chen2018} seems to be consistent with observations on the $j$-$R$ diagram, where $j$ and $R$ are the specific angular momentum of cores (angular momentum per unit of mass) and the core radius, respectively. However, \cite{Ntormousi2019} reported larger angular momentum compared with observations. The relation between the initial cloud properties and the resultant core angular momentum is not fully understood. \cite{Kuznetsova2020} shows that the direction and amplitude of the angular momentum of accreting gas onto the core both change with time, and episodic accretions do not lead to the monotonic growth of the angular momentum of the cores formed in their simulation. In spite of the intensive study by previous works, the internal structure of the core angular momentum is still unclear. Since the $j$-$R$ diagram is affected by the difference of the evolutionary stage of the cores considered in the sample (as discussed in \cite{Misugi2023}), the analysis of the internal angular momentum profile of each core at the same evolutionary stage is needed. \par 
Therefore, as described above, we should understand the relation between the physical properties of the parental filament and the resultant angular momentum of the cores formed from the filament fragmentation. In \cite{Misugi2019}, we showed, using semi-analytical models, that the angular momentum of molecular cloud cores formed along the filaments with the Kolmogorov turbulent velocity field is consistent with observations. Observations suggest that the density and velocity power spectra measured along the filament crests also follow the Kolmogorov-like power spectrum \citep{Roy2015, Arzoumanian2022}.  \cite{Misugi2023} (hereafter Paper\Rnum{1}) investigate the evolution of the angular momentum of cores in filaments using Godunov smoothed particle hydrodynamics method. However, since the magnetic fields is not included in Paper\Rnum{1}, the impact of the magnetic field on the angular momentum transfer in filamentary molecular clouds is still unclear. The aim of this paper is to study the effect of the magnetic fields on the evolution of the angular momentum of cores in filamentary molecular clouds. \par
This paper is organized as follows. We describe our numerical method and setup in Section \ref{sec:setup}. The results of our simulations are shown in Section \ref{sec:results}. In Section \ref{sec:discussion}, we present a discussion of our results. We summarize this paper in Section \ref{sec:summary}. \par

\section{Numerical Setup} \label{sec:setup}

In our simulations, we solve the following magnetohydrodynamic equations with self gravity:

\begin{eqnarray}
\label{eq:basiceqeom}
\frac{\mathrm{d} {\bm v} }{\mathrm{d} t}=-\frac{1}{\rho}\left\{ \bm{\nabla} \left(P+\frac{1}{2} B^{2}\right)- \bm{\nabla} \cdot({\bm B} {\bm B})\right\}  + \bm{\nabla} \int d x^{3} \frac{ \mathrm{G} \rho\left(x^{\prime}\right)}{\left|x-x^{\prime}\right|},
\end{eqnarray}

\begin{eqnarray}
\label{eq:basiceqind}
\frac{d}{d t}\left(\frac{\bm{B}}{\rho}\right)=\left(\frac{\bm{B}}{\rho} \cdot \bm{\nabla} \right) {\bm v},
\end{eqnarray}
where $\rho$, $P$, ${\bm B}$, and ${\bm v}$ are the gas density, gas pressure, magnetic filed, and gas velocity, respectively. $\mathrm{G}$ is the gravitational constant. $\bm{\nabla}$ is the derivative with respect to ${\bm x}$, which is the position vector. We adopt an isothermal equation of state. \par
To solve Equations \ref{eq:basiceqeom} and \ref{eq:basiceqind}, we use a Godunov smoothed particle magnetohydrodynamic (GSPM) method \cite[][Iwasaki et al. in prep.]{Iwasaki2011} with the hyperbolic divergence cleaning method for GSPM \citep{Iwasaki2013}. We adopt the Barnes-Hut tree algorithm \citep{Barnes1986} with opening angle of 0.4 to calculate the gravity. Our code is parallelized by using the Framework for Developing Particle Simulator \cite[FDPS, ][]{Iwasawa2016} to accelerate the calculation. We apply a periodic boundary condition in the $z$-direction which is parallel to the filament axis. We put $N$ copies of the filament to mimic the periodic boundary condition in the $z$-direction\footnote{To check whether our treatment of the boundary condition can reproduce the periodic boundary condition, we calculate the gravitational profile of the equilibrium filament in our simulation to compare with the analytical gravitational profile corresponding to the equilibrium density profile (Equation \ref{eq:inidenpro}). We measure the gravitational profile of the equilibrium filament by changing $N$. $N=0$  corresponds to the finite length isolated filament. For $N=0$, the gravitational profile is far from the equilibrium profile of the infinitely long filament. If we adopt $N>2$, the gravitational profile converges to the analytical profile. We use $N=4$ in this paper and confirm that the convergence is stable up to $t = 10 \ t_{\rm ff}$ when $N=4$.}. In this paper, we use $N=4$. In the $x$ and $y$-directions, we put the SPH particles in the region where the Alfv\'en wave sweeps in the computational time. This means that the Alfv\'en wave does not reach the boundary of the $x$ and $y$ directions in the computational time. The mass of a given particle in our simulations is $2.5 \times 10^{-5} \ {\rm M}_{\odot}$. The initial filament is modeled with about $2 \times 10^6$ SPH particles.\par

\begin{figure*}[]
\gridline{\fig{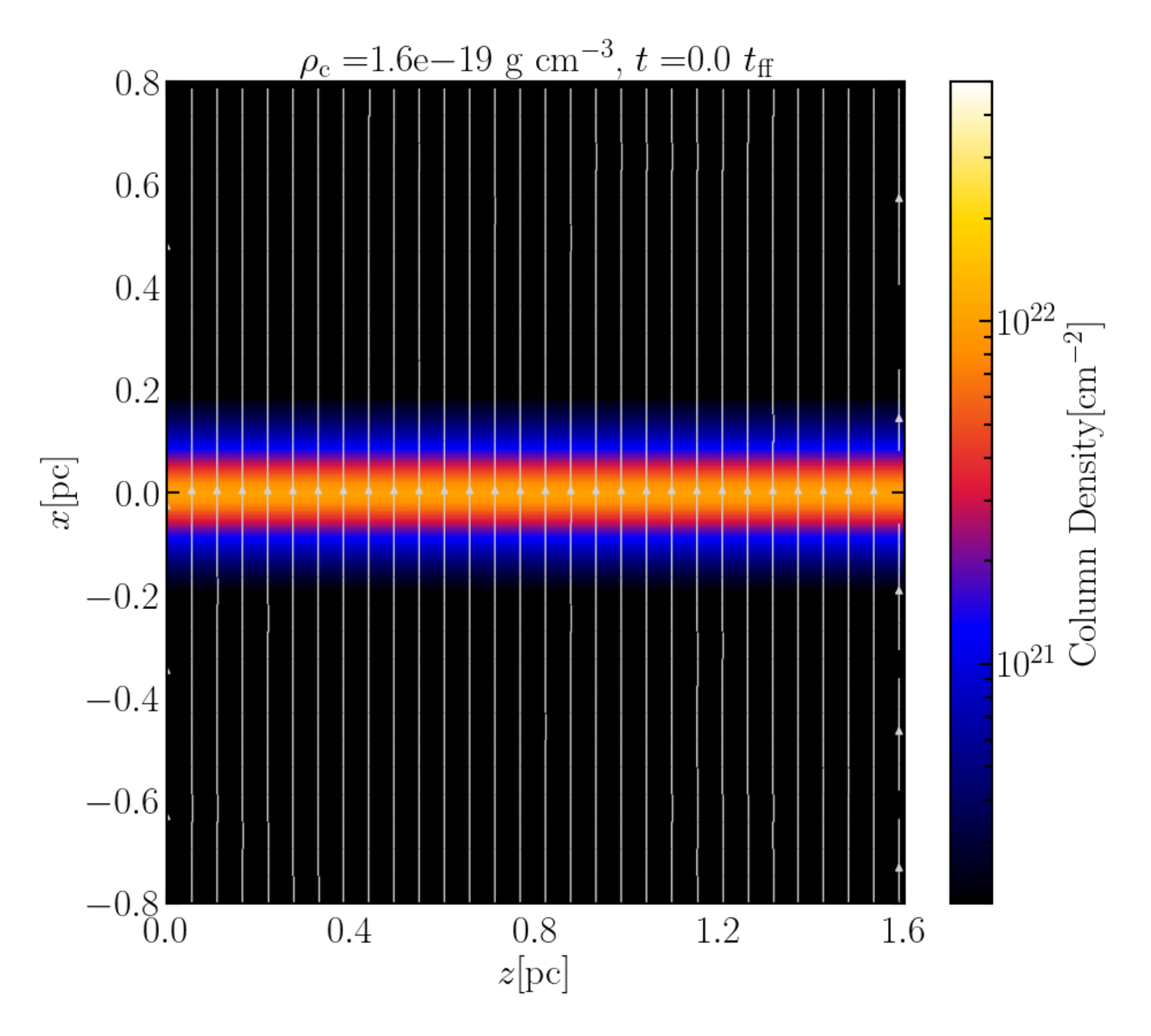}{0.37\textwidth}{(a)}
          \fig{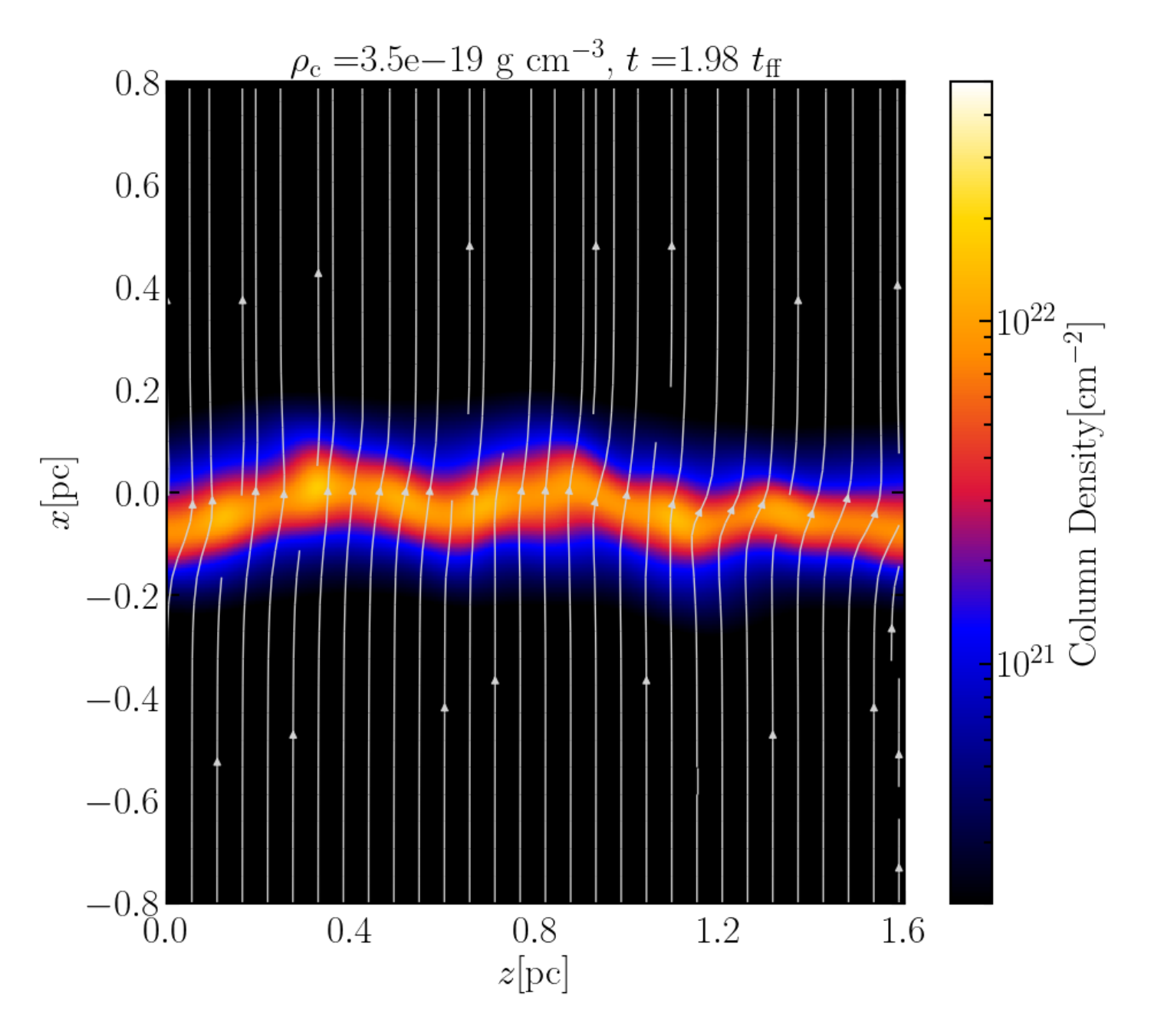}{0.37\textwidth}{(b)}
          }
\gridline{\fig{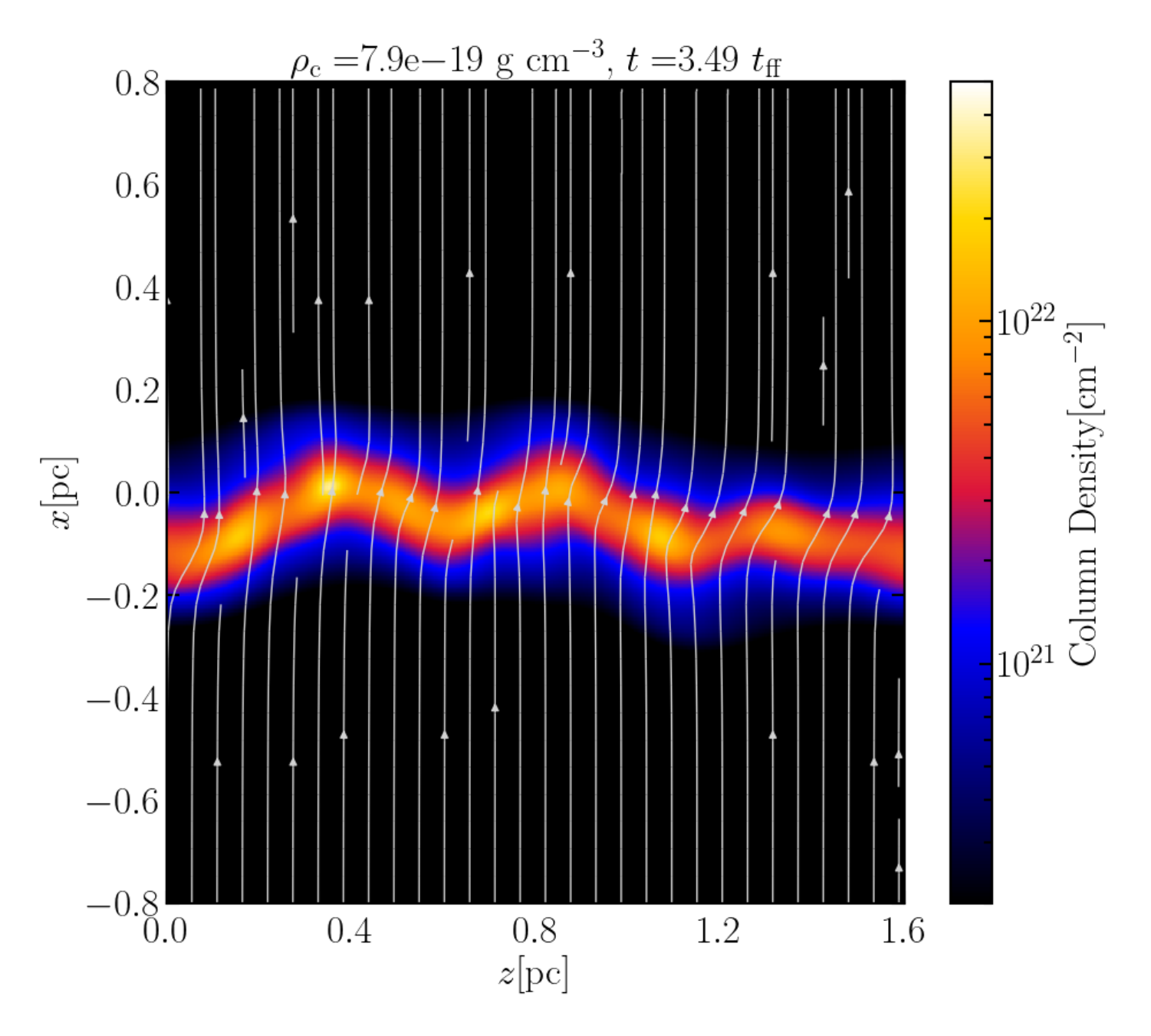}{0.37\textwidth}{(c)}
	\fig{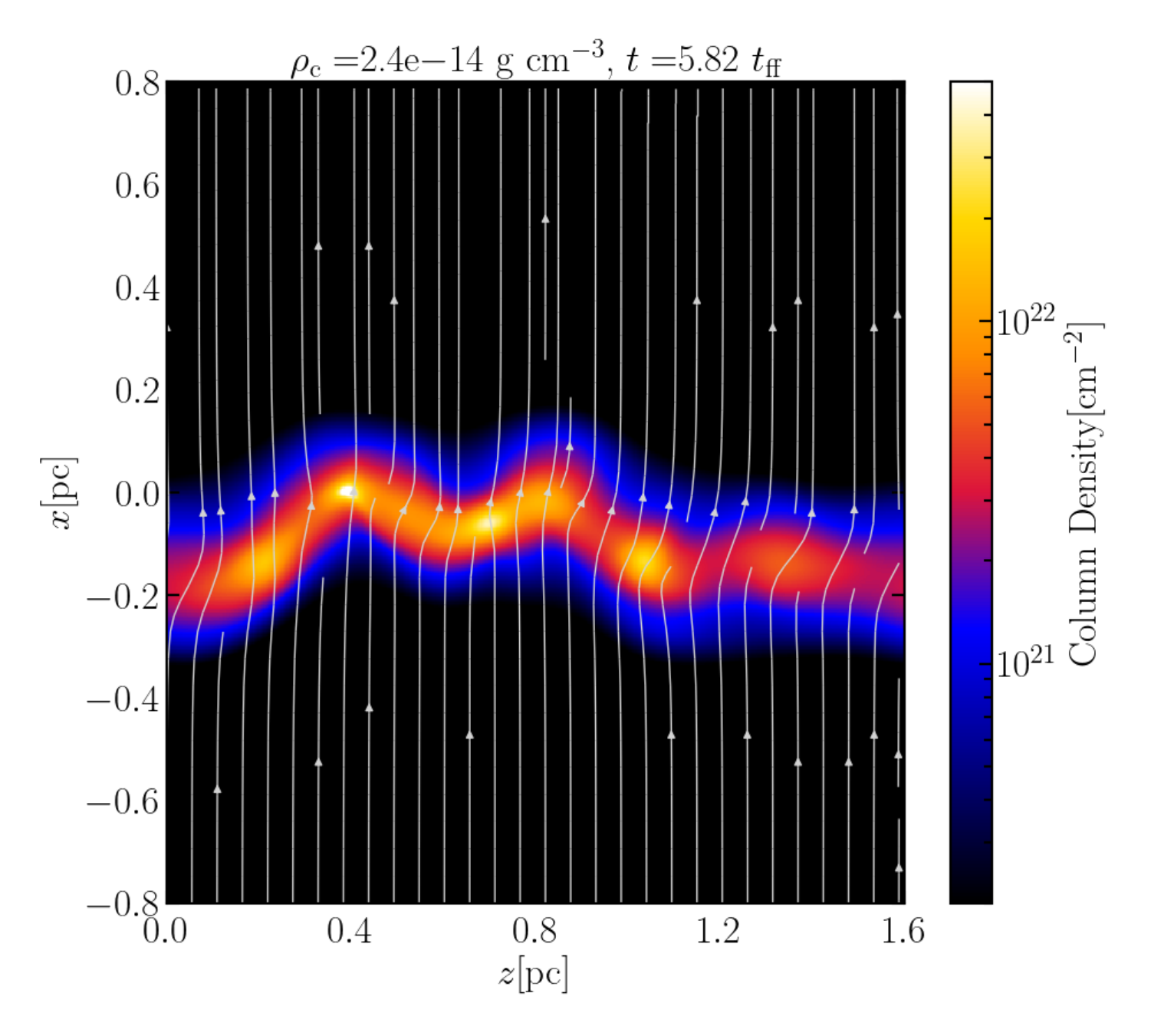}{0.37\textwidth}{(d)}
          }
\caption{Time sequence of the fragmentation of a filament threaded by perpendicular magnetic fields where the initial magnetic field has a strength of $B_0=10 \ \mu G$ (from panel (a) to (d)). The color scale represents the column density calculated by integrating the density along the $y$-axis. The elapsed time and maximum density are denoted in each panel. $t_{\rm ff}$ is the free fall time defined by $t_{\rm ff}=(4\pi G \rho_{\rm c0})^{-1/2} = 0.079 \ {\rm Myr}$. The white lines are the magnetic field lines. }
\label{fig:filevo}
\end{figure*}

The initial condition of the simulation corresponds to a filament following the hydrostatic equilibrium density profile:
\begin{eqnarray}
\label{eq:inidenpro}
\rho(r)=\rho_{\mathrm{c} 0}\left[1+\left(\frac{r}{H_{0}}\right)^{2}\right]^{-2}.
\end{eqnarray}
Here, $\rho_{\rm c0}$ is the peak density along the filament axis, $r$ is the radius in the cylindrical coordinate system, and $H_{0}$ is the scale height defined by
\begin{eqnarray}
H_{0} \equiv \sqrt{\frac{2 c_{\rm s}^{2}}{\pi {\mathrm G} \rho_{\rm c0}}},
\label{eq:scaleheight}
\end{eqnarray}
where $c_{\rm s}=0.2\  {\rm km \ s^{-1}}$ is the sound speed for $T=10 \  {\rm K}$. In this paper, we adopt $H_{0} = 0.05\ {\rm pc}$ since observations suggest that the filaments have a universal width of 0.1 pc \citep{Arzoumanian_2011,Arzoumanian2019}. The line mass of the filament is equal to the critical line mass, ${\rm M}_{\rm line,crit}=18 \ {\rm M_{\odot} pc^{-1}}$ for $T=10$ K, implying a peak density of $3.4 \times 10^4 {\rm cm}^{-3}$. We numerically generate the turbulent velocity field along the filament following the method described in Paper\Rnum{1}. In this paper, we adopt the Kolmogorov turbulence as the initial turbulent velocity field. The initial three-dimensional velocity dispersion is $\sigma=2c_{\rm s}$ for our fiducial model. Note that the transonic velocity fluctuations are measured in recent observations for filaments with line masses close to the critical value \citep{Hacar2011,Arzoumanian2013,Hacar2016}. The initial magnetic field is along the $x$-axis and perpendicular to the filament ($z$-) axis. We construct two models with the initial magnetic field strength of $B_0=2 \ {\rm \mu G}$ and $B_0=10 \ {\rm \mu G}$ to investigate the effect of the initial magnetic field on the resultant angular momentum of cores. The initial plasma beta at the ridge of the filament in the case of $B_0=2 \ {\rm \mu G}$ and $B_0=10 \ {\rm \mu G}$ are $\beta = 400$ and $16$, respectively. This means that $B_0=2 \ {\rm \mu G}$ is compatible with the hydrodynamic case shown in Paper\Rnum{1}. $B_0=10 \ {\rm \mu G}$ is still weaker than the observed average value at the volume density $n_{\rm c}\sim 10^{4} \ {\rm cm^{-3}}$ \citep{Pattle2022}. However, since the observed strength of the magnetic field is widely scattered, it is meaningful to investigate the evolution of the filament in the case of $B_0=10 \ {\rm \mu G}$. Note that the strength of the magnetic field of the cores is amplified by the fragmentation of the filament and subsequent contraction of the cores. The importance of the magnetic field compared with gravity is often described using the mass-to-flux ratio \citep{Tomisaka1988}. The non-dimensional mass-to-flux ratio in the case of $B_0=2 \ {\rm \mu G}$ and $B_0=10 \ {\rm \mu G}$ are $\mu = 2\pi G^{1/2}\Sigma/B_0 = 31.4$ and $6.28$, respectively. $\Sigma$ is the column density at the ridge of the filament. This means that all cores formed in our simulations are magnetically supercritical. We perform 40 sets of simulations for the weak magnetic field case ($B_0=2 \ {\rm \mu G}$) and 40 sets of simulations for the strong magnetic field case ($B_0=10 \ {\rm \mu G}$) using different turbulent seeds in each parameter to study the evolution of the core angular momentum statistically. The initial turbulent seed is the same between the weak magnetic field case and the strong magnetic field case. We run each simulation until the maximum density reaches $\rho_{\rm crit}=2.8\times 10^{-14} \ {\rm g \ cm^{-3}} = 2\times 10^5\rho_{\rm c0} $ ($n_{\rm crit}=7.3\times 10^{9} \ {\rm cm^{-3}}$). Here, $\rho_{\rm c0}=1.4\times 10^{-19} \ {\rm g \ cm^{-3}}$ ($n_{\rm c0}=3.4\times 10^{4} \ {\rm cm^{-3}}$) is the initial peak density of the filament.

\section{Results} \label{sec:results}

\subsection{Overview} \label{subsec:overview}

Figure \ref{fig:filevo} displays the fragmentation of a filament in our simulation. The cores are formed along the filament due to the growth of initial turbulent fluctuations. The magnetic field lines are perpendicular to the filament axis at the initial state and are deformed by the turbulence and the core formation motions as time progresses. These deformations impact also the overall shape of the filament, which is straight at the initial stage. Since the density profile rapidly drops in the outer region of the filament ($\rho \propto r^{-4}$) and plasma beta is so small, the magnetic field lines are bent only close to the ridge of the filament. The perpendicular magnetic fields with the free boundary condition cannot halt the fragmentation of the filaments \citep{Hanawa2017}.

\subsection{Internal Structure of Core Rotation \label{subsec:internalAM}}

\begin{figure*}
\gridline{\fig{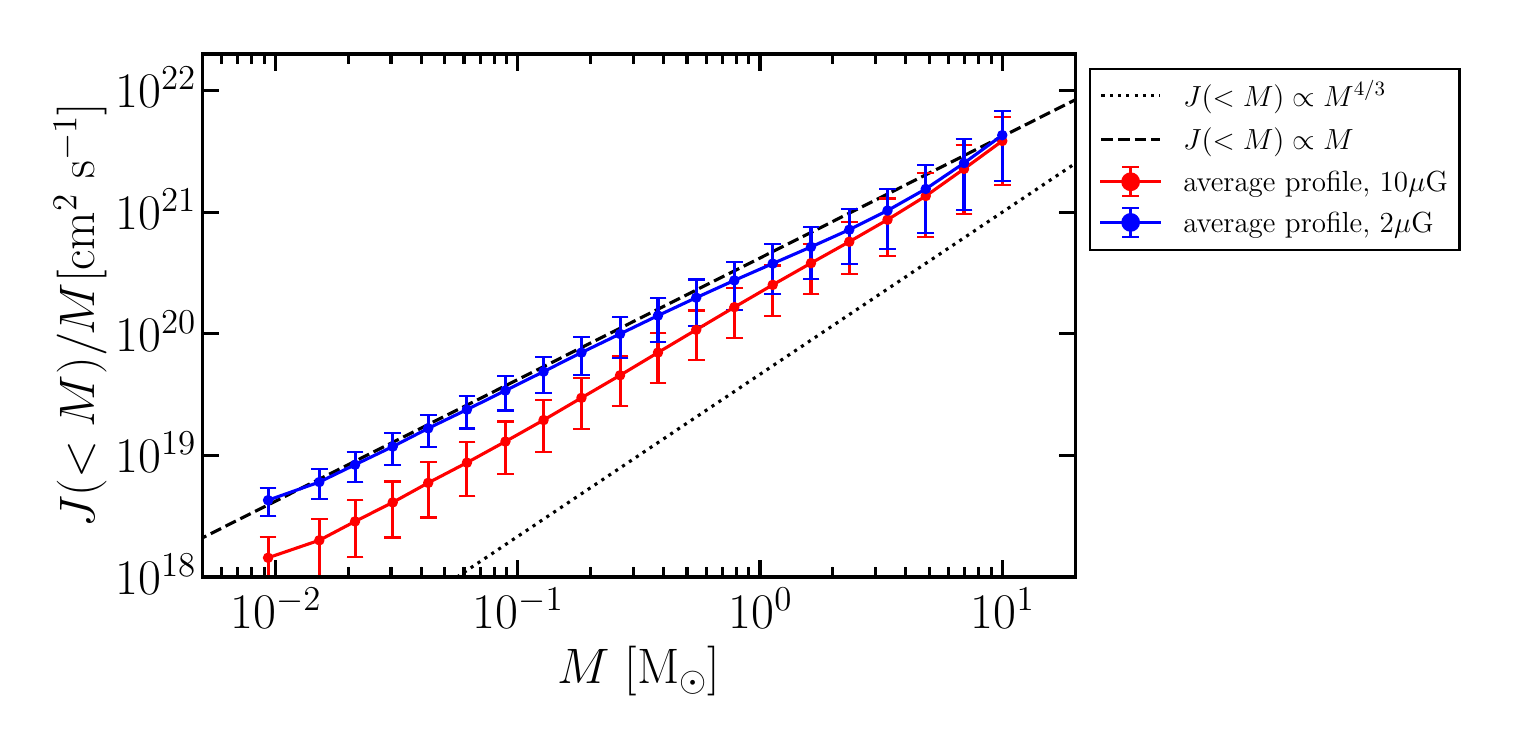}{0.7\textwidth}{(a)}
          }
\gridline{\fig{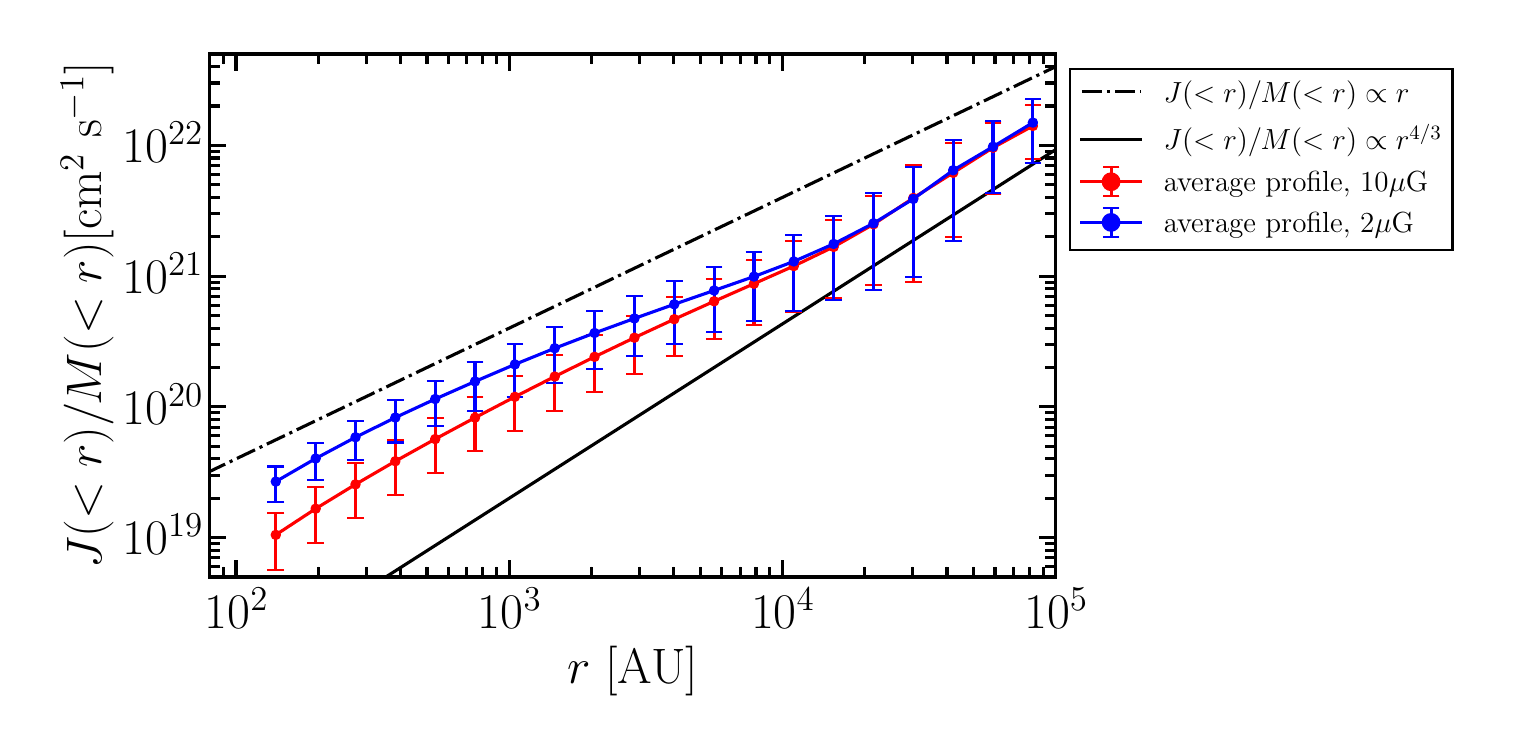}{0.7\textwidth}{(b)}
          }
\caption{(a) Averaged $j$-$M$ profile for the 40 cores. The red and blue solid lines represent the averaged $j$-$M$ profile with $B_0=10\ {\rm \mu G}$ and $B_0=2\ {\rm \mu G}$, respectively. The black dotted and dashed lines are $j \propto M^{4/3}$ and $j \propto M$, respectively.  The error bars represent the standard deviation. (b) Averaged $j$-$r$ profile for the 40 cores. The black dashed-dotted and solid lines are $j \propto r$ and $j \propto r^{4/3}$, respectively.  \label{fig:jMavecompfinal} }
\end{figure*}

Here we study the internal rotational structure of the cores averaging the results of the 40 simulations for both sets of simulations with weak and strong magnetic field. Figure \ref{fig:jMavecompfinal} (a) displays the averaged $j$-$M$ profile in the cores at the final state of our simulations, where $j$ and $M$ are the specific angular momentum and the mass of the spherical enclosed region around the density peak, respectively. The black dashed line represents $j \propto M$, which corresponds to the self-similar profile \citep{Basu1997,Saigo1998,Misugi2023}. Figure \ref{fig:jMavecompfinal} (a) indicates that the profile of the angular momentum converges to the self similar solution in the weak magnetic field case similarly to the hydrodynamic case as shown in Paper\Rnum{1}. On the other hand, in the strong magnetic field case, the angular momentum of the inner region is smaller and the slope of the profile is steeper than that of the weak magnetic field case. This result suggests that the transfer of the angular momentum is more efficient in the strong magnetic field case. Note that the dependence of the angular momentum on the strength of the initial magnetic field can be seen only in the inner region $M \lesssim 3.0 \ {\rm M}_{\odot}$ which corresponds to the radii $r \lesssim  0.1 \ {\rm pc}$. In the outer part of the core, for enclosed mass $M \gtrsim 3.0 \ {\rm M}_{\odot}$, the angular momentum profiles are similar for both weak and strong magnetic field cases. The black dotted line in Figure \ref{fig:jMavecompfinal} (a) represents the angular momentum profile due to the turbulence. Since the initial turbulent velocity field follows the Kolmogorov turbulence,  the relation between the velocity dispersion $\sigma$ and the length scale $l$ is $\sigma \propto l^{1/3} $. Therefore, the specific angular momentum can be described as $j \sim \sigma l \propto l^{4/3}$. If the length scale is larger than the width of the filament ($l \gg 2H_0$), the mass contained in the spherical regions follows $M \sim M_{\rm line} l$. Using these relations, we can derive $j \sim \sigma l \propto M^{4/3}$. The angular momentum profile of the outer region shown in Figure \ref{fig:jMavecompfinal} (a) is consistent with $j  \propto M^{4/3}$ both in the weak and the strong magnetic field cases. This indicates that the angular momentum transfer due to the magnetic field is inefficient in the outer region ($M \gtrsim 3.0 \ {\rm M}_{\odot}$). When the radius of the spherical region is larger than the width of the filament, the derived physical properties of the region is affected not only by the core but also by the filament. These results suggest that, although the magnetic breaking is not important at the filament scale, magnetic breaking is important for the angular momentum transfer in the core. \par
The averaged $j$-$r$ profile for the 40 cores at the final state is shown in Figure \ref{fig:jMavecompfinal} (b), here $r$ denotes the radial distance from the density peak of the core. The resultant angular momentum in the outer region ($\gtrsim 10^4 \ {\rm AU}$) does not depend on the strength of the initial magnetic field similarly to the $j$-$M$ relation and follows $j \propto r^{4/3}$. In the inner region ($r \lesssim 10^3 \ {\rm AU} $), the resultant profile with the weak magnetic field at the initial state follows the self similar profile ($j \propto r$). In the weak magnetic field case, a shallower slope appears for $10^3\ {\rm AU} \lesssim r \lesssim 10^4 \ {\rm AU} $. This is due to the contraction during the core formation phase in the filaments. The displacement of the fluid elements due to the contraction at $r \sim 10^3 \ {\rm AU}$ is larger than that at $r \sim 10^4 \ {\rm AU}$, consequently, a shallower slope appears at $10^3\ {\rm AU} \lesssim r \lesssim 10^4 \ {\rm AU} $ (compare to the $j \propto r^{4/3}$ at $r \gtrsim 10^4 \ {\rm AU}$) before the transition to the $ j \propto r $ in the inner part of the core at $r \lesssim 10^3 \ {\rm AU}$. In the strong magnetic field case, the region with shallower slope is not clearly seen since the angular momentum transfer is efficient due to the strong magnetic field. We will analyze the angular momentum transfer in more detail in the following subsections.

\subsection{Evolution of the Angular Momentum of the Core \label{subsec:evototalAM}}

\begin{figure*}
\epsscale{0.9}
\plotone{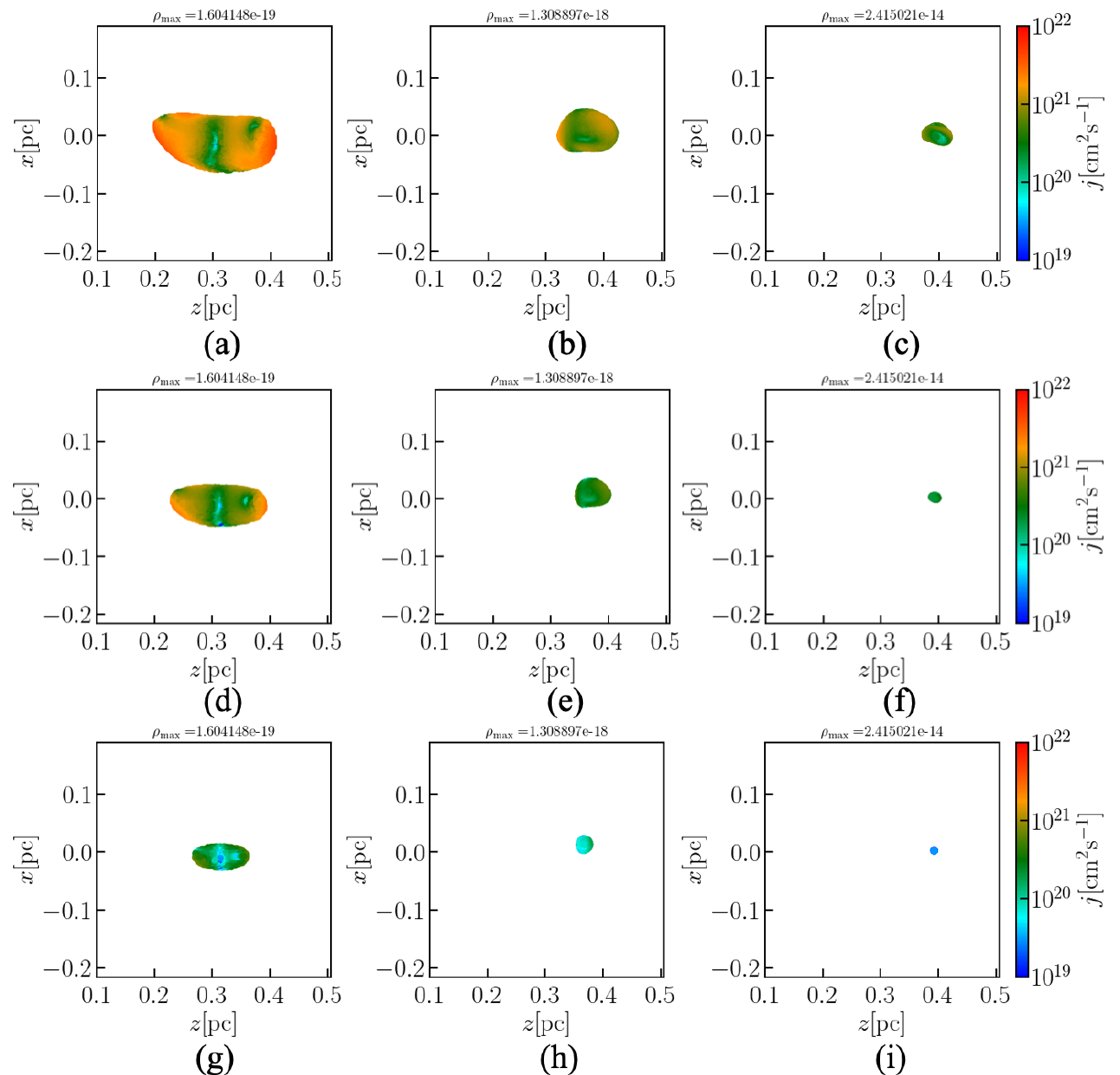}
\caption{Time sequence of the morphology of the core on the $x$-$z$ plane from left to right. The color represents the specific angular momentum of the SPH particles around the center of mass of the core. The top, middle, and bottom rows correspond to the evolution of the core morphology with the enclosed mass of $M_{\rm ana}=3.0$, $1.0$, and $0.1 \ {\rm M}_{\odot}$, respectively. $M_{\rm ana}$ is the enclosed mass in the density contour at the final state of our simulations. \label{fig:lagtrace} }
\end{figure*}

\begin{figure}[t]
\begin{tabular}{cc}
\begin{minipage}[t]{.35\textwidth}
\centering
\includegraphics[width=7cm]{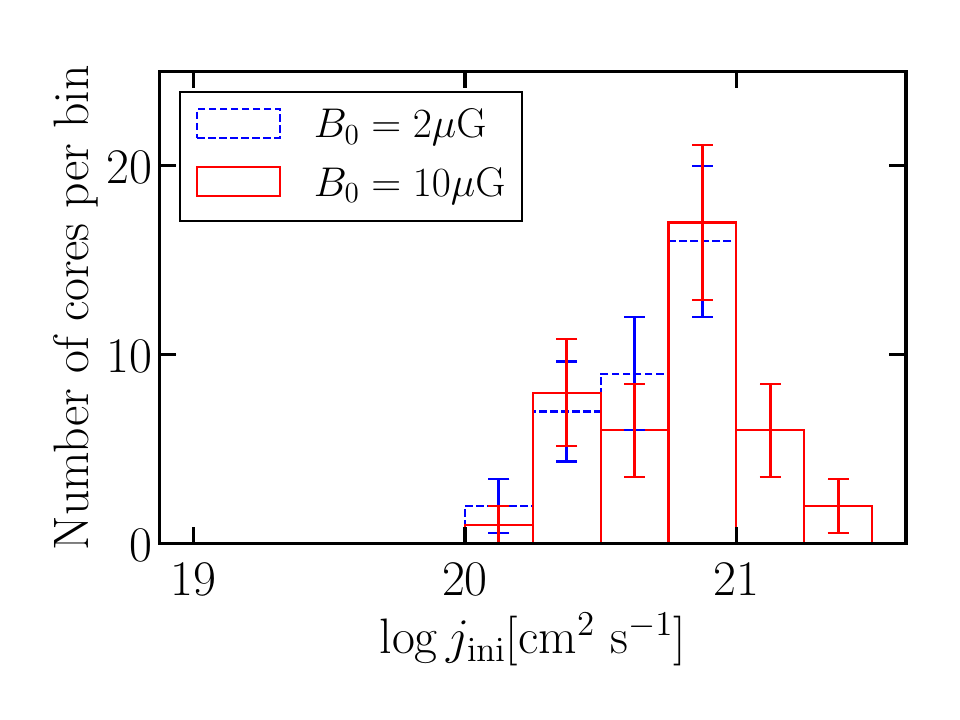}
\end{minipage}
\begin{minipage}{.15\textwidth}
\hspace{10mm}
\end{minipage}

\begin{minipage}[t]{.35\textwidth}
\centering
\includegraphics[width=7cm]{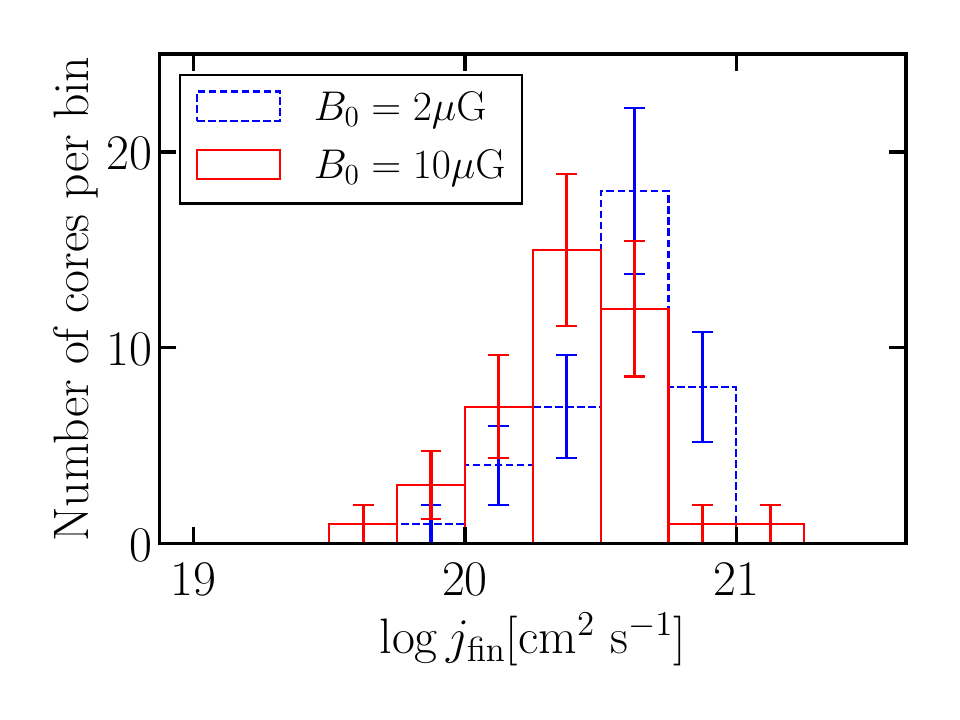}
\end{minipage}
\end{tabular}
\caption{Histograms of the specific angular momentum of all the cores at the initial state (left) and at the final state (right). The blue dashed and red solid histograms are the results with the strength of the initial magnetic field of $B_0=10\ {\rm \mu G}$ and $B_0=2\ {\rm \mu G}$, respectively. The error bars correspond to statistical uncertainties. \label{fig:coream_hist} }

\end{figure}

\begin{figure}
\epsscale{0.7}
\plotone{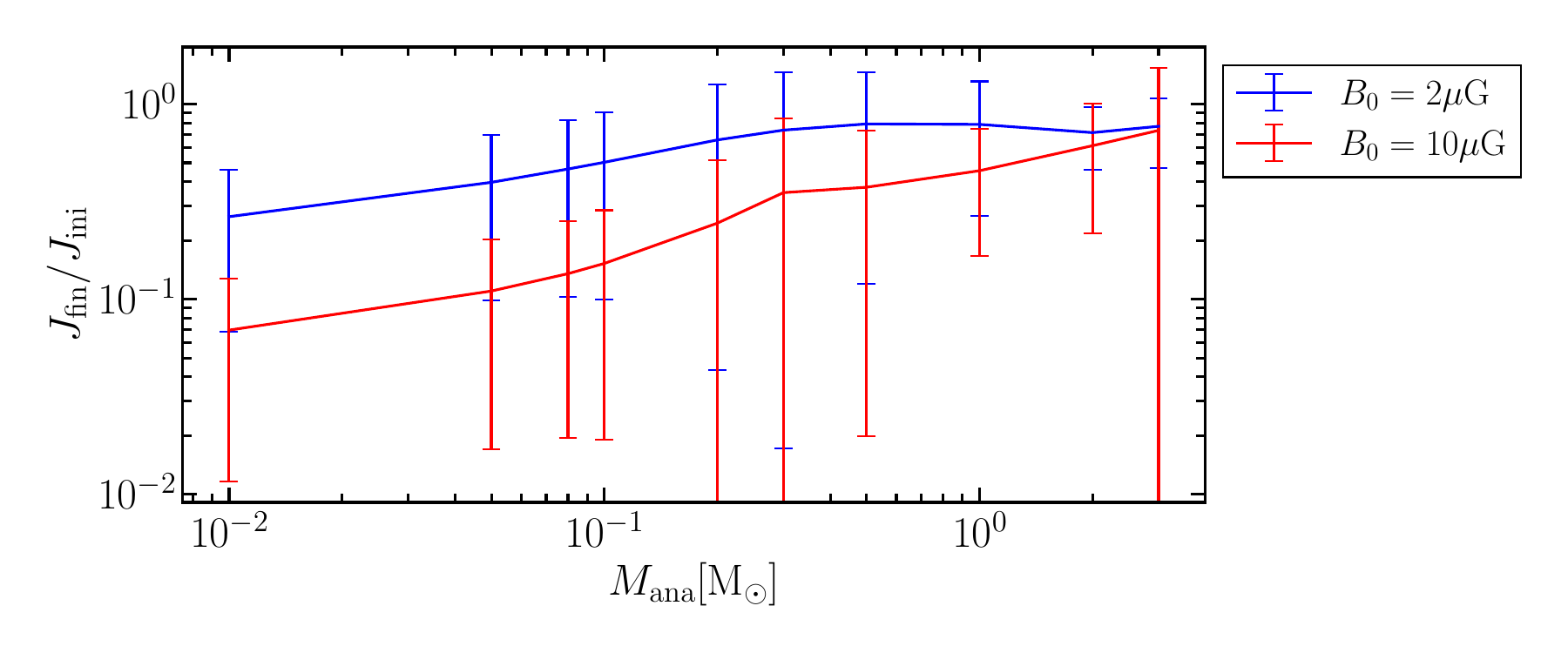}
\caption{Ratio of the angular momentum at the final state to that at the initial state averaged over the 40 cores at different mass scale. The red and blue solid lines are in the case of $B_0=10\ {\rm \mu G}$ and $B_0=2\ {\rm \mu G}$. The horizontal axis is $M_{\rm ana}$ defined as the enclosed mass at the final state (Equation \ref{eq:Mana}). \label{fig:spejratio} }
\end{figure}

In Section \ref{subsec:internalAM}, we discussed the internal angular momentum profile at the final state. In this subsection, we investigate the evolution of the core angular momentum by tracing the trajectory of the SPH particles. To do so, first, we identify the core around the density peak along the filament at the final state of the simulation. That core corresponds to the fastest collapsing core along the filament. Then, we determine the mass scale $M_{\rm ana}$ which we analyze. We choose a density contour threshold $\rho_{\rm ana}$ to satisfy the following equation:
\begin{eqnarray}
M_{\rm ana}( \rho_{\rm ana} ) = \int_{\rho > \rho_{\rm ana}} \rho d^3x.
\label{eq:Mana}
\end{eqnarray}
Note that the enclosed region is not spherical since the region is defined by the density contour in contrast with the previous subsection (Section \ref{subsec:internalAM}). In this analysis, $M_{\rm ana}$ is in the range of $0.01 \ {\rm M}_{\odot}$ to  $3.0 \ {\rm M}_{\odot}$. We take the periodic boundary conditions into account. Then, the trajectories of the SPH particles are traced from the final state to the initial state of the simulation. Figure \ref{fig:lagtrace} displays the evolution of the SPH particles which form the core at the final state for $M_{\rm ana}=3.0 \ {\rm M}_{\odot}$, $1.0 \ {\rm M}_{\odot}$, and $0.1 \ {\rm M}_{\odot}$. At the initial state, the core is elongated along the $z$-axis which is parallel to the longitudinal axis of the filament. The aspect ratio decreases with time due to the contraction along the filament axis (Figure \ref{fig:lagtrace}). \par

We calculate the angular momentum of the core around the center of mass $\boldsymbol{J}$ at each mass scale as follows:
\begin{eqnarray}
\boldsymbol{J}=\sum_{i} m_i ( \boldsymbol{x}_i-\boldsymbol{x}_{\rm c}) \times (\boldsymbol{v}_i-\boldsymbol{v}_{\rm c}),
\label{eq:angcal}
\end{eqnarray}
where $\boldsymbol{x}_i$ and $\boldsymbol{v}_i$ denote the position and velocity vectors of the SPH particles, respectively. $\boldsymbol{x}_{\rm c}$ and $\boldsymbol{v}_{\rm c}$ are the position and velocity of the center of mass of the core, respectively. The summation includes the SPH particles which belong to the core at the final state. The specific angular momentum of a core is defined as follows:
\begin{eqnarray}
\boldsymbol{j}=\frac{|\boldsymbol{J}|}{M_{\rm ana}}.
\label{eq:samdef}
\end{eqnarray}
We note that, if there is no interaction with the ambient gas around the core, the angular momentum should be conserved. In our simulations, since the cores are formed and embedded in the filament, they interact with the surrounding medium. The histograms of the specific angular momentum of the cores at the initial and at the final state are displayed in Figure \ref{fig:coream_hist}. The left panel of Figure \ref{fig:coream_hist} shows that the distribution of the initial angular momentum of the cores does not depend on the strength of the magnetic field. The mean of the distribution at the final state with $B_0=2 \ {\rm \mu G}$ and $B_0=10 \ {\rm \mu G}$ are $ j = 5.0 \times 10^{20} \ {\rm cm^2 \ s^{-1}}$ and $ j = 3.6 \times 10^{20} \ {\rm cm^2 \ s^{-1}}$, respectively (Figure \ref{fig:coream_hist} right), suggesting a more efficient transport of the angular momentum in the strong magnetic field case. Figure \ref{fig:spejratio} shows the dependence of the ratio of the angular momentum at the final state to that at the initial state on the mass scale. Although the angular momentum transfer is efficient in small mass scales ($M_{\rm ana} \lesssim 0.1 \ {\rm M_{\odot} }$), The average ratio $J_{\rm fin}/J_{\rm ini}$ at $M_{\rm ana}=1.0 \ {\rm M_{\odot} }$ in the case of weak and strong magnetic field case are 0.78 and 0.46, respectively. These results indicate that, although the resultant angular momentum of the cores with the strong magnetic field is smaller than that obtained in the weak magnetic field case, the difference of the angular momentum between both cases is relatively small at $M_{\rm ana} \gtrsim1.0 \ {\rm M_{\odot} }$. To investigate which force dominates the angular momentum transfer at each mass scale, we will analyze the evolution of the torques in the next subsection (Section \ref{subsec:AMtransfer}).

\subsection{Mechanism of Angular Momentum Transfer \label{subsec:AMtransfer}}
\begin{figure*}
\gridline{\fig{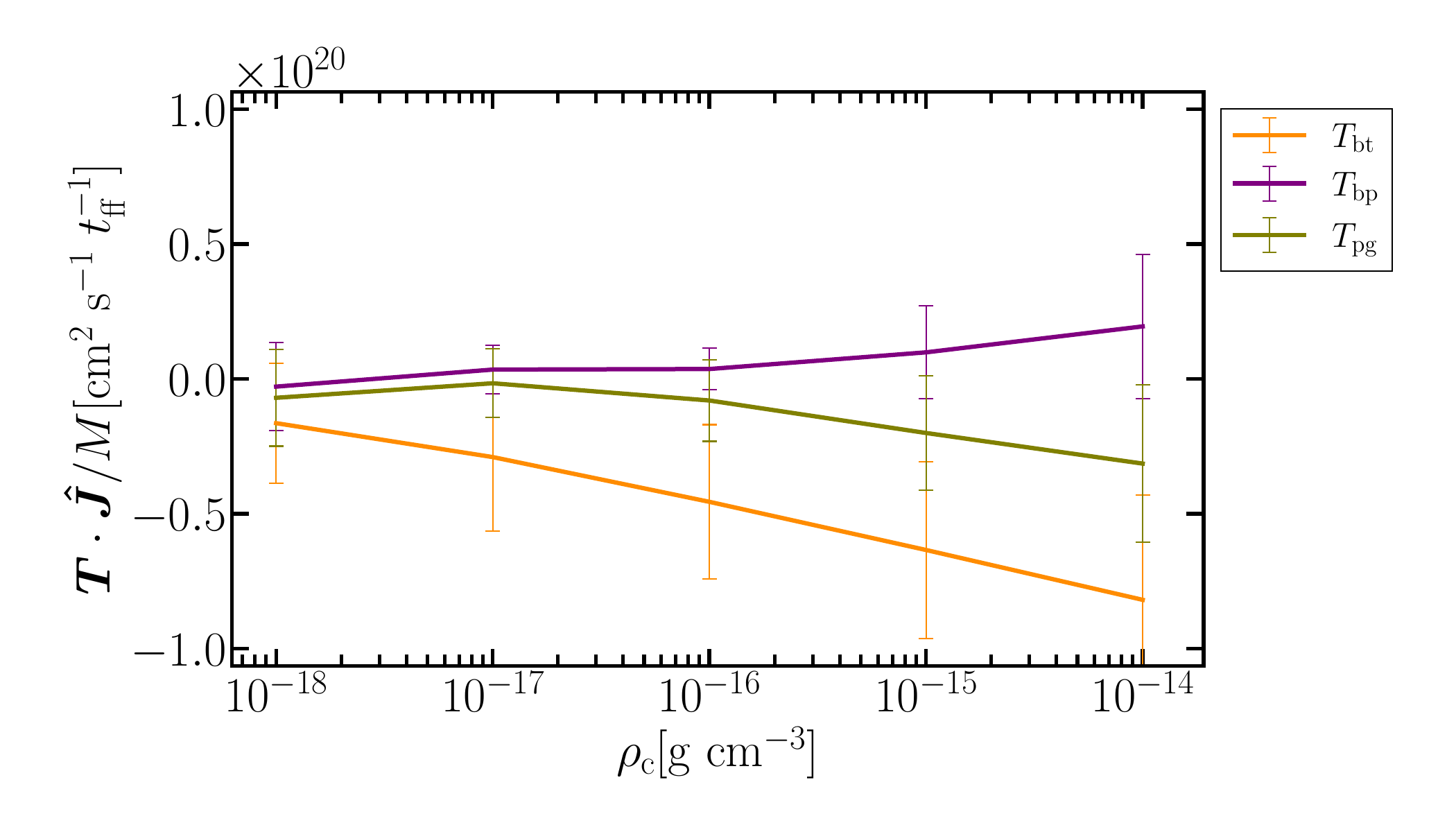}{0.45\textwidth}{(a) $0.1 \ {\rm M}_{\odot}$}
          \fig{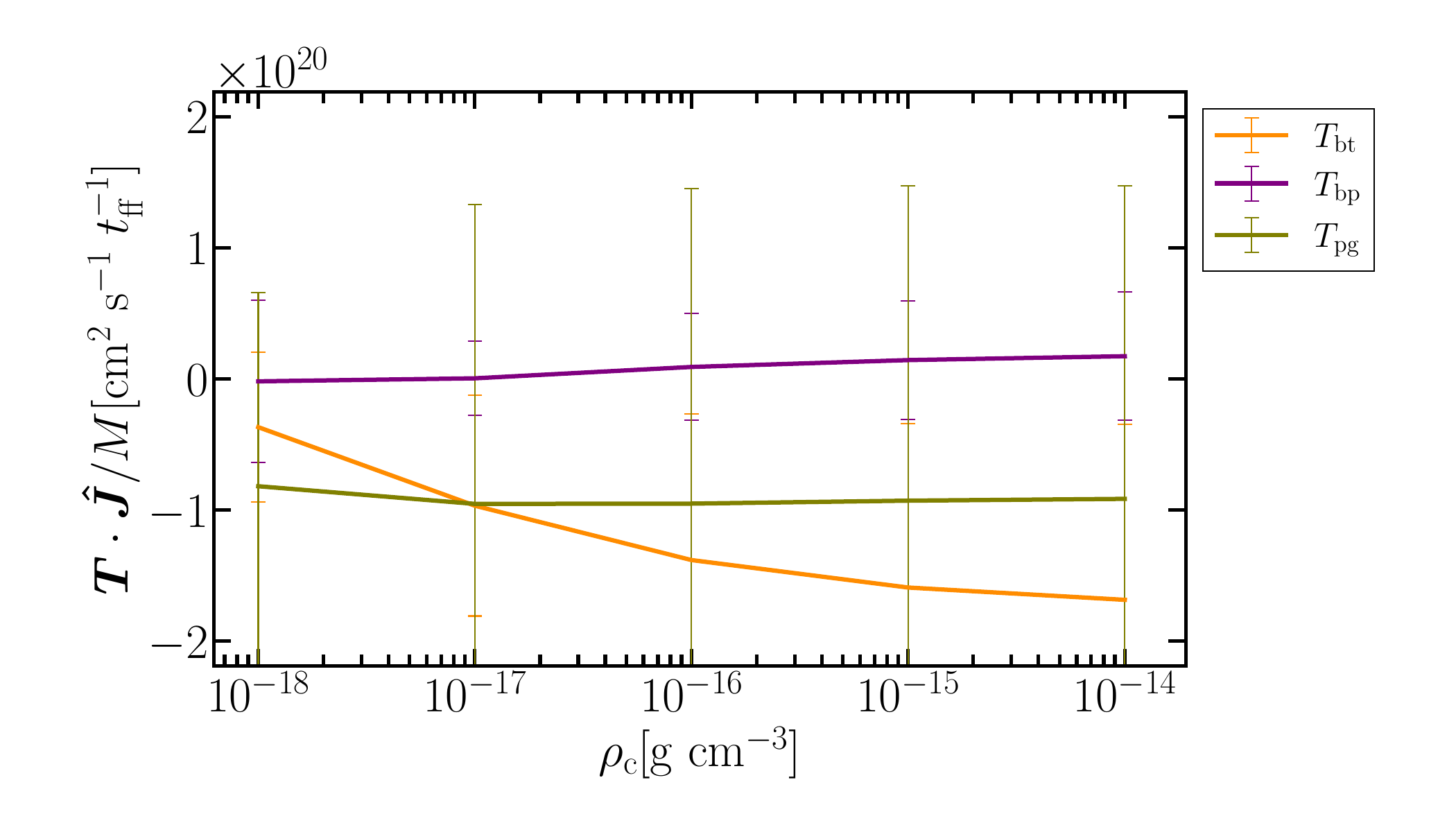}{0.45\textwidth}{(b) $1.0 \ {\rm M}_{\odot}$}
          }
\gridline{\fig{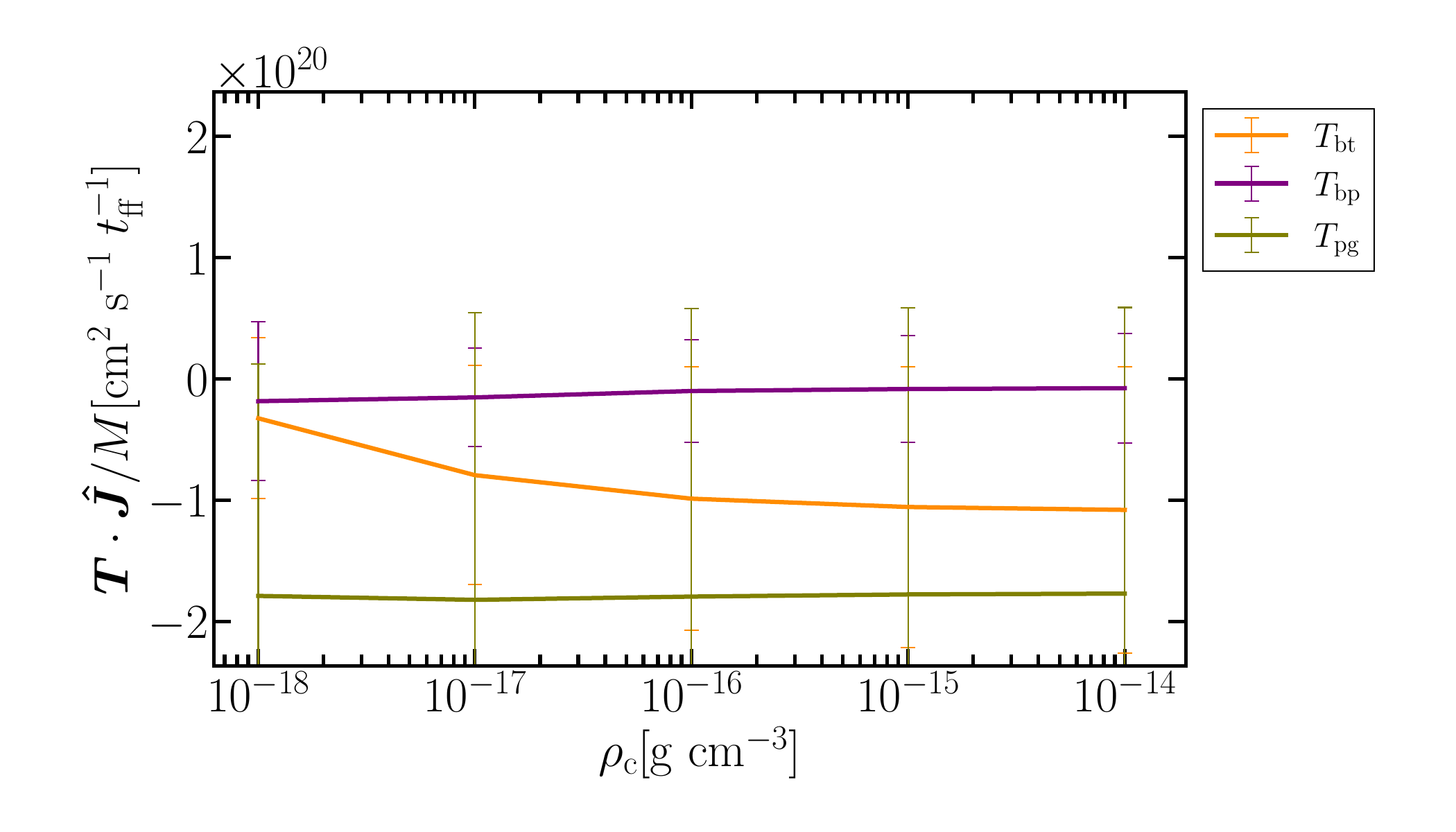}{0.45\textwidth}{(c) $3.0 \ {\rm M}_{\odot}$}
          }
\caption{Torque exerted on the cores of (a) $0.1 \ {\rm M}_{\odot}$, (b) $1.0 \ {\rm M}_{\odot}$, and (c) $3.0 \ {\rm M}_{\odot}$, respectively, in the strong magnetic field case. The torques are averaged over 40 cores. The vertical axis is an inner product between the torque and the unit vector of the angular momentum. The horizontal axis shows the central density and hence the time evolution of the core. If the inner product is negative, the angular momentum is extracted. If the inner product is positive, the core gains the angular momentum. The orange, magenta, and olive solid lines are the magnetic tension, magnetic pressure, and the sum of the gas pressure and gravity torques, respectively. }
\label{fig:torque}
\end{figure*}

To investigate the mechanism of angular momentum transfer in the cores, we calculate the torque that is acting on the core. The torque $ \boldsymbol{T}$ can be written formally as follows:
\begin{eqnarray}
\label{eq:torqueformally}
\boldsymbol{T}=\int \rho \left(\boldsymbol{x}-\boldsymbol{x}_{\rm c}\right) \times \boldsymbol{F}d^3x,
\end{eqnarray}
where $\boldsymbol{F}$ is a force per unit of mass. 
We discretize Equation \ref{eq:torqueformally} and calculate the magnetic tension torque ($\boldsymbol{T}_{\mathrm{bt}}$), magnetic pressure torque ($\boldsymbol{T}_{\mathrm{bp}}$), gas pressure torque ($\boldsymbol{T}_{\mathrm{f}}$), and gravity torque ($\boldsymbol{T}_{\mathrm{g}}$) as follows:

\begin{eqnarray}
\label{eq:tenstorque}
\boldsymbol{T}_{\mathrm{bt}}=\sum_{i} m_i \left(\boldsymbol{x}_{i}-\boldsymbol{x}_{\rm c}\right) \times \boldsymbol{F}_{\mathrm{bt}, i}, 
\end{eqnarray}

\begin{eqnarray}
\label{eq:bptorque}
\boldsymbol{T}_{\mathrm{bp}}=\sum_{i} m_i \left(\boldsymbol{x}_{i}-\boldsymbol{x}_{\rm c}\right) \times \boldsymbol{F}_{\mathrm{bp}, i}, 
\end{eqnarray}

\begin{eqnarray}
\label{eq:gastorque}
\boldsymbol{T}_{\mathrm{f}}=\sum_{i} m_i \left(\boldsymbol{x}_{i}-\boldsymbol{x}_{\rm c}\right) \times \boldsymbol{F}_{\mathrm{f}, i}, 
\end{eqnarray}

\begin{eqnarray}
\label{eq:gratorque}
\boldsymbol{T}_{\mathrm{g}}=\sum_{i} m_i \left(\boldsymbol{x}_{i}-\boldsymbol{x}_{\rm c}\right) \times \boldsymbol{F}_{\mathrm{g}, i}, 
\end{eqnarray}
where $\boldsymbol{F}_{\mathrm{bt}, i}$, $\boldsymbol{F}_{\mathrm{bp}, i}$, $\boldsymbol{F}_{\mathrm{f}, i}$, and $\boldsymbol{F}_{\mathrm{g}, i}$ are the magnetic tension force, the magnetic pressure force, the gas pressure force, and the gravity per unit of mass of the $i$th particle.\footnote{Here the notation is different from Paper\Rnum{1}. In Paper\Rnum{1}, the torque $\boldsymbol{T}$ and the force $\boldsymbol{F}$ are defined as the torque per mass and the force per volume, respectively.  } In the case of the weak magnetic field, the mechanism of angular momentum transfer is the same as that without the magnetic field as show in Paper\Rnum{1}. In the hydrodynamic case, the angular momentum of the core is transferred by the negative pressure torque during the core formation phase from the filament fragmentation ($\rho_{\rm c}< 10^{-17}\ {\rm g \ cm^{-3}}$). This can be understood as follows. First, most of the cores rotate perpendicularly to the filament axis, and the major axis of the core is inclined with respect to the filament longitudinal axis at the initial state. Then, as time progresses, the major axis of the cores aligns with the filament longitudinal axis with the rotation axis of the cores perpendicular to the filament axis. This rotational direction makes it easier for the cores to gather their mass. Since, at the initial state, the filament follows the hydrostatic equilibrium profile, the pressure torque and the gravitational torque are negative and positive, respectively, with respect to the core rotation axis which is perpendicular to the filament axis. However, in the runaway collapse phase, the effect of the initial filament geometry decreases. Since the core has a non-axisymmetric structure and its inner region rotates faster than the outer region, both torques become negative.\par
Figure \ref{fig:torque} shows the evolution of the torques exerted on the cores in the strong magnetic field case. The orange, magenta, and olive solid lines represent the magnetic tension, magnetic pressure, and the sum of the gas pressure and gravity torques, respectively. Figure \ref{fig:torque} shows that the magnetic pressure term does not contribute to the angular momentum transfer for all mass scales. As shown in Figure \ref{fig:torque} (a), the magnetic braking is the dominant mechanism to transfer the angular momentum at $0.1 \ {\rm M}_{\odot}$. Since the magnetic field lines are twisted due to the spin up of the inner region of the core, the magnetic tension increases as time progresses. The magnetic pressure and the sum of the gas pressure and gravity torques do not play an important role in the angular momentum transfer compared to the magnetic braking at the mass scale of $0.1 \ {\rm M}_{\odot}$.  At $1.0 \ {\rm M}_{\odot}$ (Figure \ref{fig:torque} (b)), although the difference between the gas pressure and gravity is responsible for the angular momentum transfer until the central density reaches $\rho_{\rm c}=10^{-17} \ {\rm g \ cm^{-3}}$, the magnetic braking becomes more important for densities larger than $\rho_{\rm c}=10^{-17} \ {\rm g \ cm^{-3}}$. On the other hand, the sum of the gas pressure and gravity torques is always larger than the magnetic tension torque at $3.0 \ {\rm M}_{\odot}$. For the small mass scales ($M_{\rm crit} \lesssim 1.0 \ {\rm M}_{\odot}$), the magnetic braking is efficient since the surrounding gas is massive enough to extract the angular momentum from the inner region. On the other hand, the mass of the surrounding gas is relatively small for the large mass scales  ($M_{\rm crit} \gtrsim 3.0 \ {\rm M}_{\odot}$), since the filament has a steep density profile in the outer region ($\rho \propto r^{-4}$). This is the reason why the magnetic braking is not efficient at $3.0  \ {\rm M}_{\odot}$. All the torques remain constant $\rho_{\rm c} \gtrsim 10^{-16} \ {\rm g \ cm^{-3}}$ because the timescale of the gravitational collapse becomes shorter and shorter as the gravitational collapse proceeds. Figure \ref{fig:torque} indicates that we can define the critical mass which determines whether the angular momentum transfer is efficient or not. If the mass scale is smaller than the critical mass, the magnetic field plays a role in the angular momentum transfer. On the other hand, if the mass scale is larger than the critical mass, the magnetic tension does not affect the evolution of the angular momentum. Our results suggest that the critical mass $M_{\rm crit}$ is $1.0 \ {\rm M}_{\odot} \lesssim M_{\rm crit} \lesssim 3.0  \ {\rm M}_{\odot}$ in the case of the strong magnetic field, because we see a change of behavior of the importance of the torques between these two mass scales (Figure \ref{fig:torque}). This behavior of torques is consistent with the results shown in Figure \ref{fig:jMavecompfinal}. Since the magnetic tension is not dominant in the outer region ($\gtrsim 3.0 \ {\rm M}_{\odot}$), the angular momentum profile of the outer region is almost the same irrespective of the strength of the initial magnetic field.

\begin{figure*}
\gridline{\fig{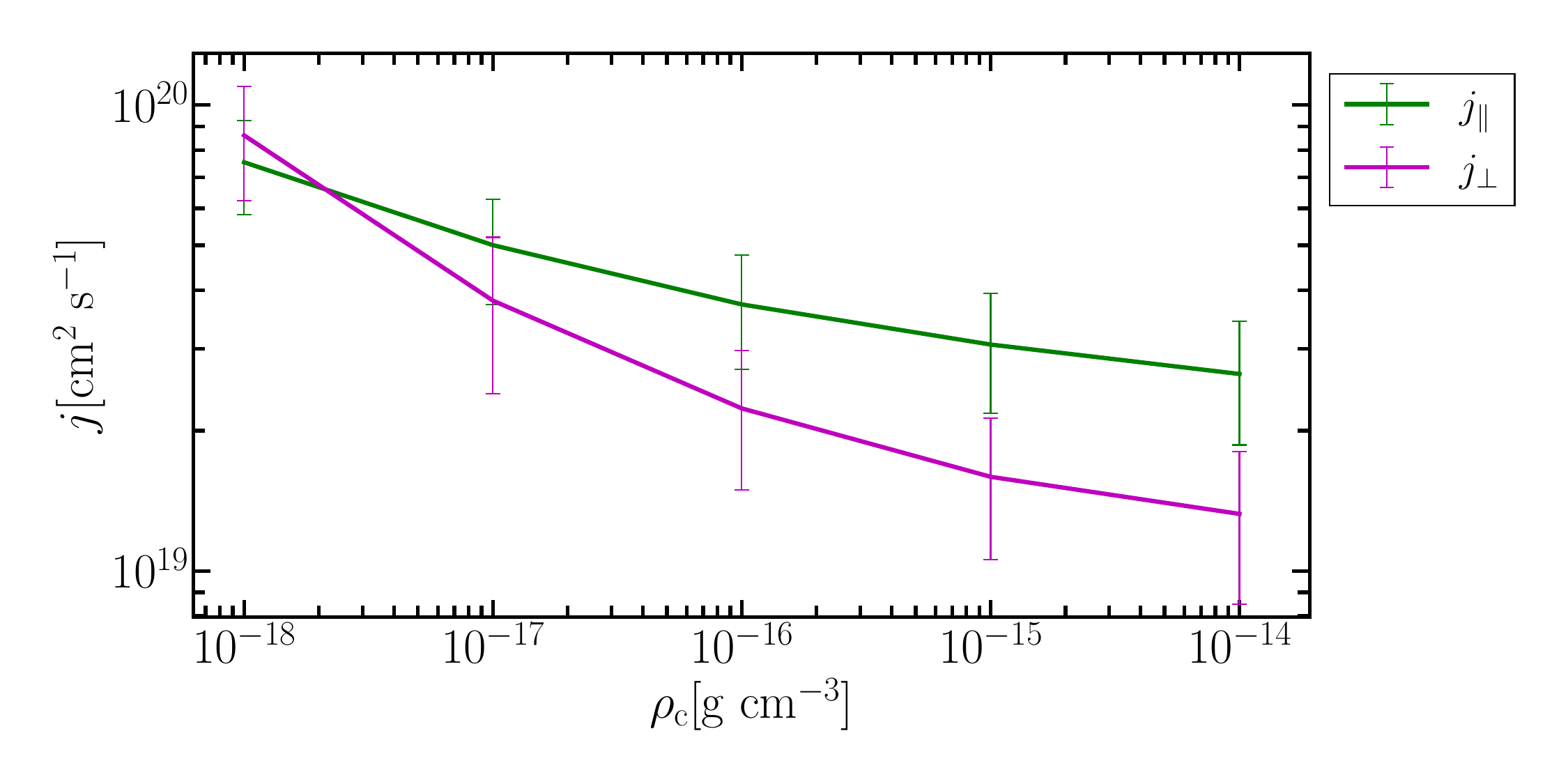}{0.50\textwidth}{(a) $0.1 \ {\rm M}_{\odot}$}
          \fig{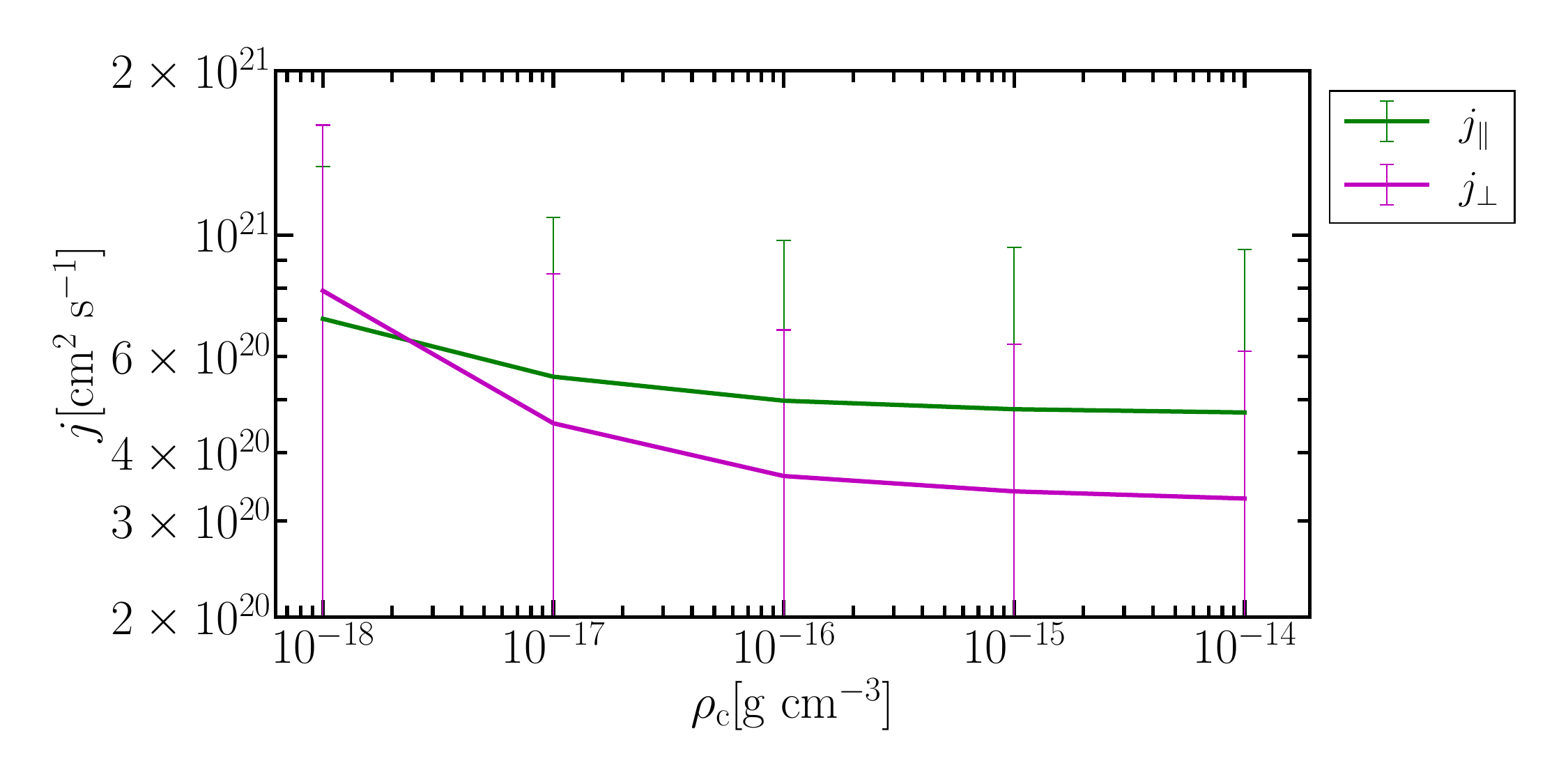}{0.50\textwidth}{(b) $1.0 \ {\rm M}_{\odot}$}
          }
\gridline{\fig{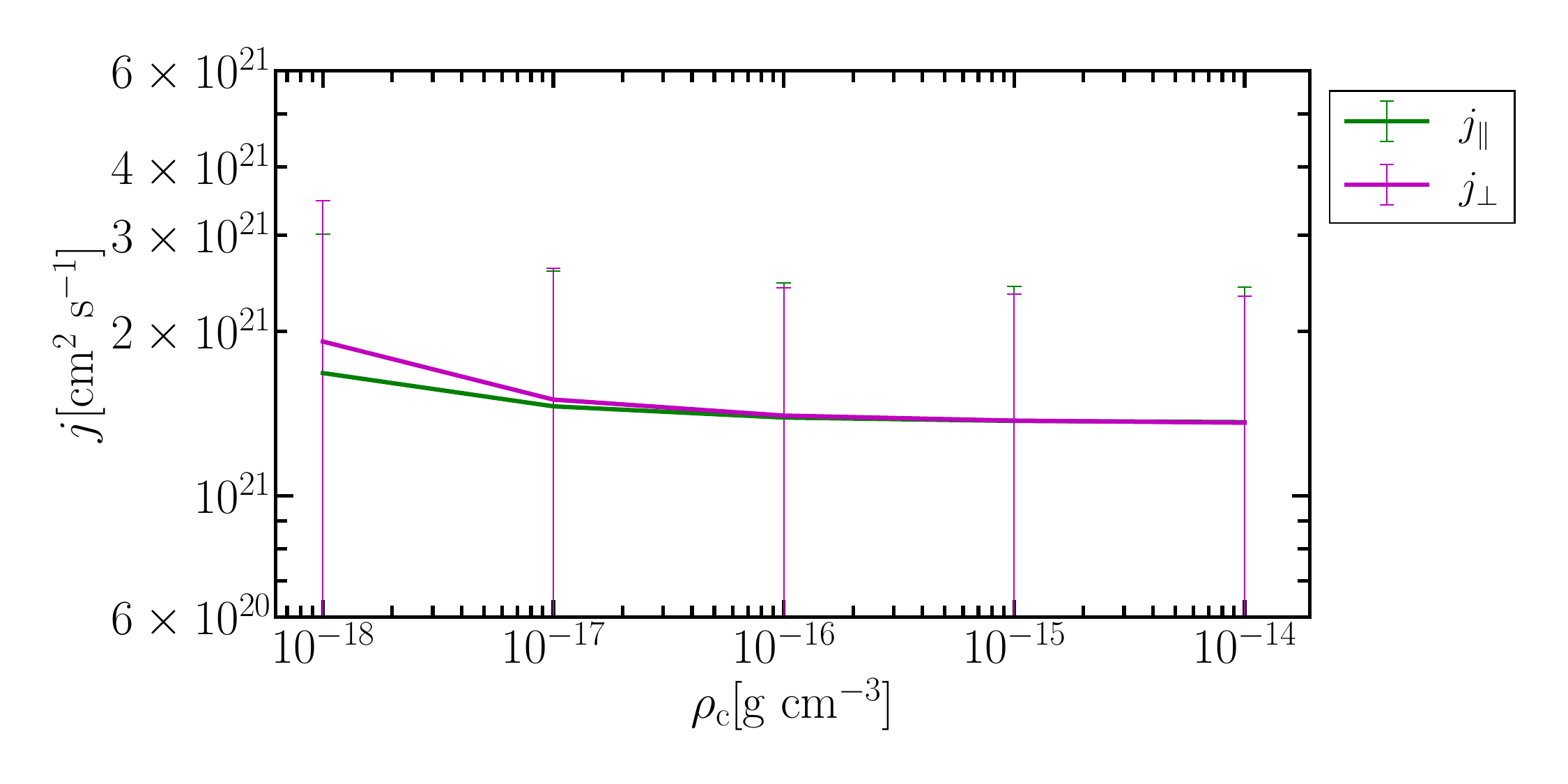}{0.50\textwidth}{(c)  $3.0 \ {\rm M}_{\odot}$}
          }
\caption{Evolution of the angular momentum of the cores with the enclosed mass of (a) $0.1 \ {\rm M}_{\odot}$, (b) $1.0 \ {\rm M}_{\odot}$, and (c) $3.0 \ {\rm M}_{\odot}$, respectively, in the strong magnetic field case. The green and purple solid lines are the specific angular momentum averaged in parallel and perpendicular samples, respectively. }
\label{fig:aniso_j_evo}
\end{figure*}

\subsection{Anisotropy of the Angular Momentum Transfer \label{subsec:anisoAMtransfer}}

To study the anisotropy of the angular momentum transfer, we divide all cores into two samples at $\rho_{\rm c}=10^{-18} \ {\rm g \ cm^{-3}}$. The first sample is composed of the cores with the angle between their angular momentum and the local magnetic field smaller than $30^{\circ}$. The cores of the second sample have the angle between their angular momentum and the local magnetic field larger than $60^{\circ}$. The former and the latter samples are referred to as the parallel sample and the perpendicular sample, respectively. The parallel sample contains 8, 9, and 10 cores at $M_{\rm ana} = 0.1,\ 1.0,\ 3.0 \ {\rm M}_{\odot}$, respectively. The perpendicular sample consists of 10, 9, and 11 cores at $M_{\rm ana} = 0.1,\ 1.0,\ 3.0 \ {\rm M}_{\odot}$, respectively. The local magnetic field is defined as the volume averaged magnetic field in the enclosed region at each time step. Note that the magnetic filed orientation can change due to the initial turbulent velocity field. Figure \ref{fig:aniso_j_evo} displays the evolution of the angular momentum of the parallel and the perpendicular samples in the strong magnetic field case. At $M_{\rm ana} = 0.1 \ {\rm M}_{\odot}$, the angular momentum of the perpendicular sample is smaller than that of the parallel sample at the final state, while both has almost the same angular momentum at $\rho_{\rm c}=10^{-18} \ {\rm g \ cm^{-3}}$. The same tendency can be seen in the case of $M_{\rm ana} = 1.0 \ {\rm M}_{\odot}$, although the difference of the angular momentum between the perpendicular and parallel sample is a bit smaller compared with the results of $M_{\rm ana} = 0.1 \ {\rm M}_{\odot}$. On the other hand, at $M_{\rm ana} = 3.0 \ {\rm M}_{\odot}$, the evolution of the angular momentum of the perpendicular sample is almost the same with that of the parallel sample. This dependence on the mass scale can be understood from the results shown in Figure \ref{fig:torque}. In Figure \ref{fig:torque}, we show that the the magnetic tension is the dominant process of the angular momentum transfer only for $M_{\rm ana} \lesssim 3.0  \ {\rm M}_{\odot}$. Since only the magnetic field causes the anisotropy in the angular momentum transfer, the difference of the angular momentum between samples can be seen only at $M_{\rm ana} = 0.1,\ 1.0 \ {\rm M}_{\odot}$.

\begin{figure*}
\gridline{\fig{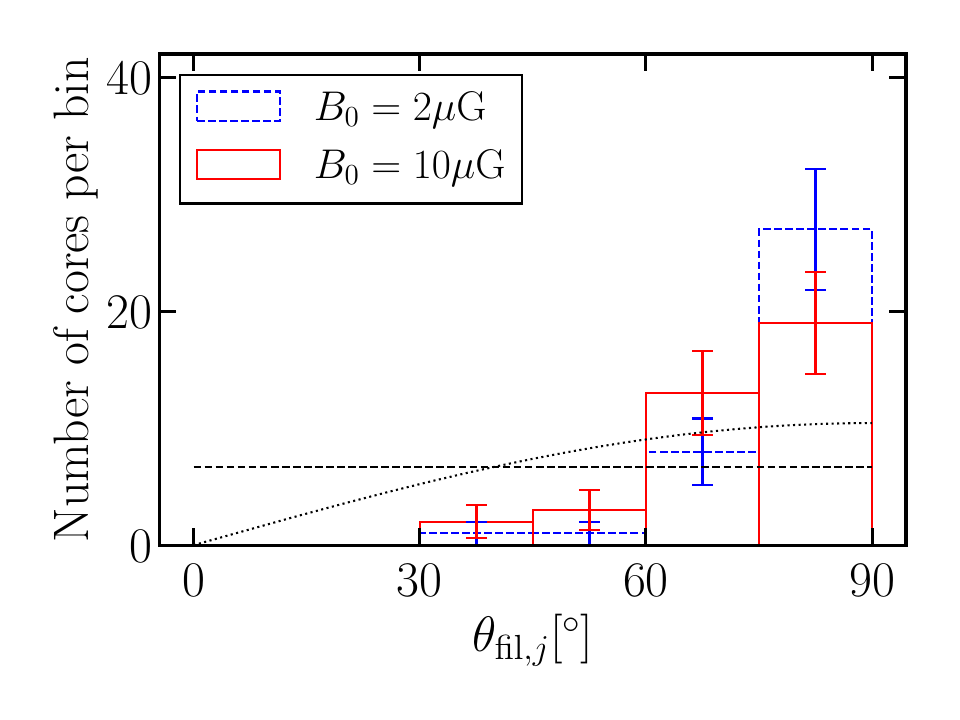}{0.50\textwidth}{(a)  $0.1 \ {\rm M}_{\odot}$}
          \fig{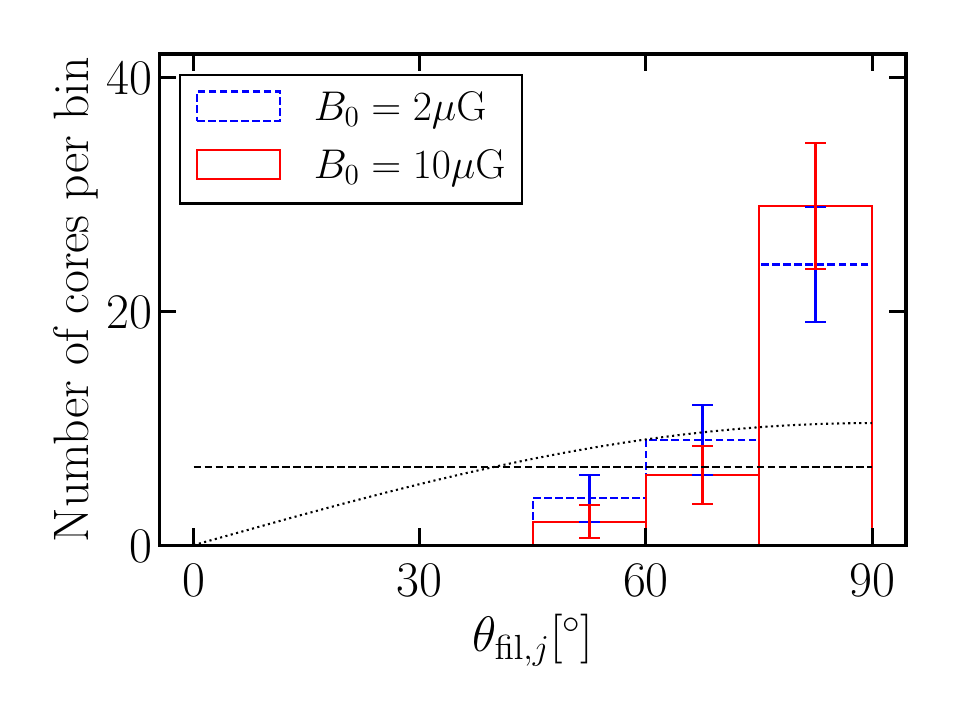}{0.50\textwidth}{(b) $1.0 \ {\rm M}_{\odot}$}
          }
\gridline{\fig{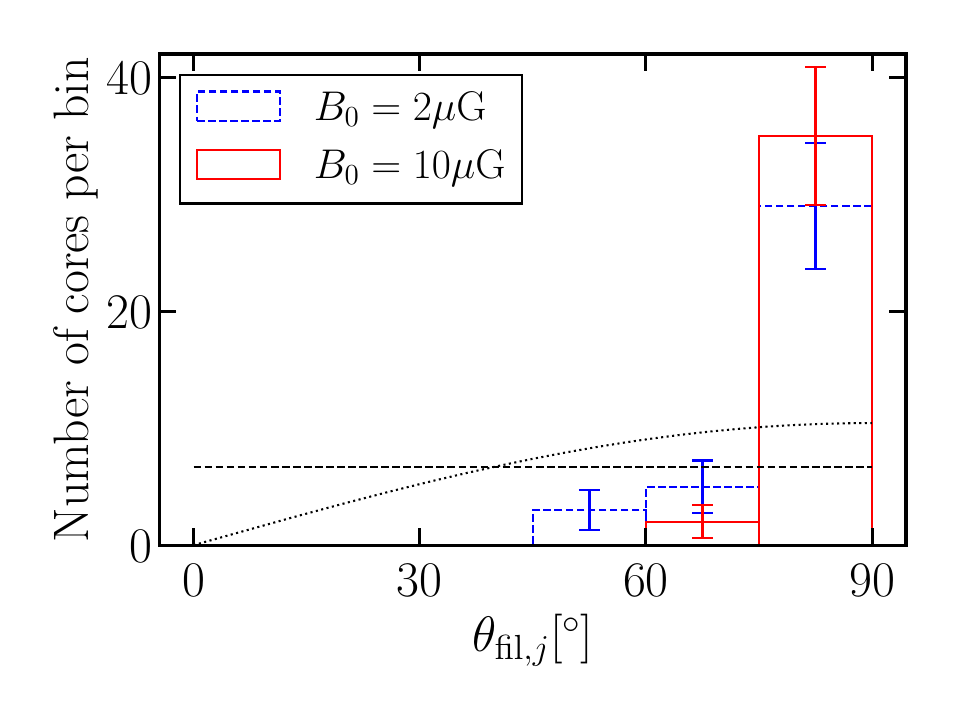}{0.50\textwidth}{(c) $3.0 \ {\rm M}_{\odot}$}
          }
\caption{Histogram of the angle between the angular momentum vector of the cores and the filament axis at the final state with (a) $0.1 \ {\rm M}_{\odot}$, (b) $1.0 \ {\rm M}_{\odot}$, and (c) $3.0 \ {\rm M}_{\odot}$. The blue dashed and red solid lines represent the case of weak and strong magnetic field cases, respectively. The black dotted line corresponds to the random distribution in the three-dimensional space taking into account the effect of the solid angle. The black dashed line represents the random distribution in the 2D plane perpendicular to the filament axis. }
\label{fig:thetajfil_hist}
\end{figure*}

\subsection{Misalignment between the Angular Momentum and the Magnetic Field \label{subsec:misalignment}}

\begin{figure*}
\gridline{\fig{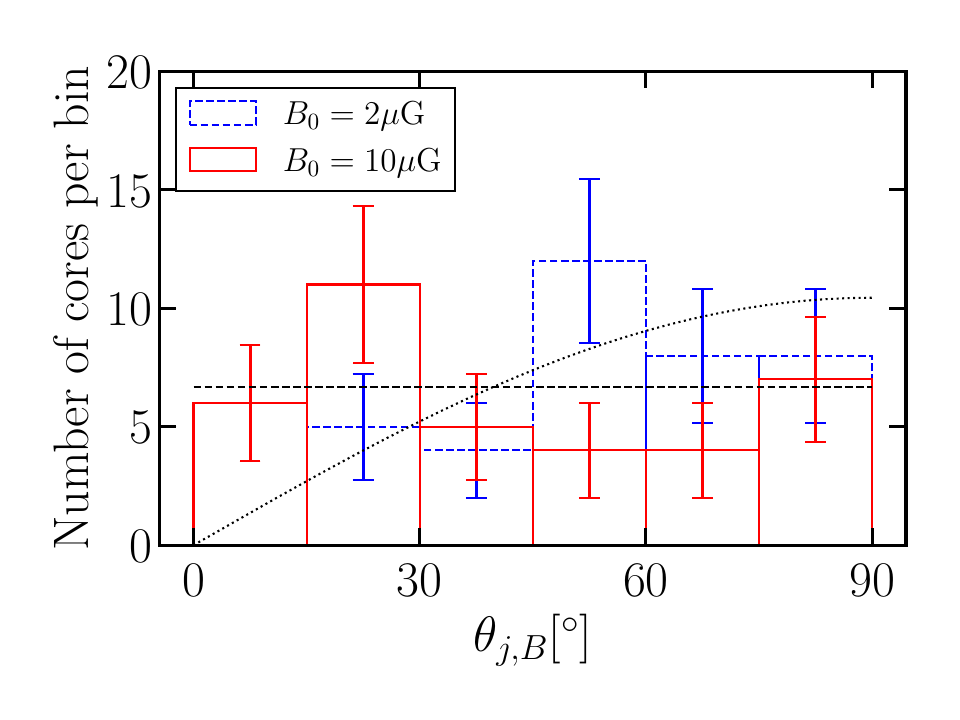}{0.50\textwidth}{(a)  $0.1 \ {\rm M}_{\odot}$}
          \fig{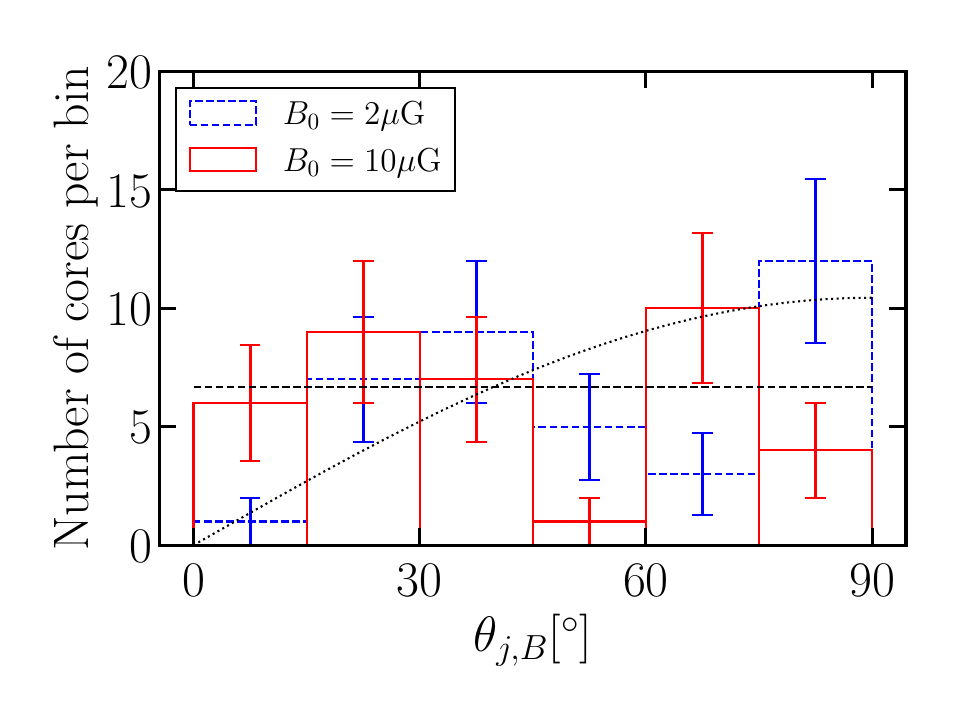}{0.50\textwidth}{(b) $1.0 \ {\rm M}_{\odot}$}
          }
\gridline{\fig{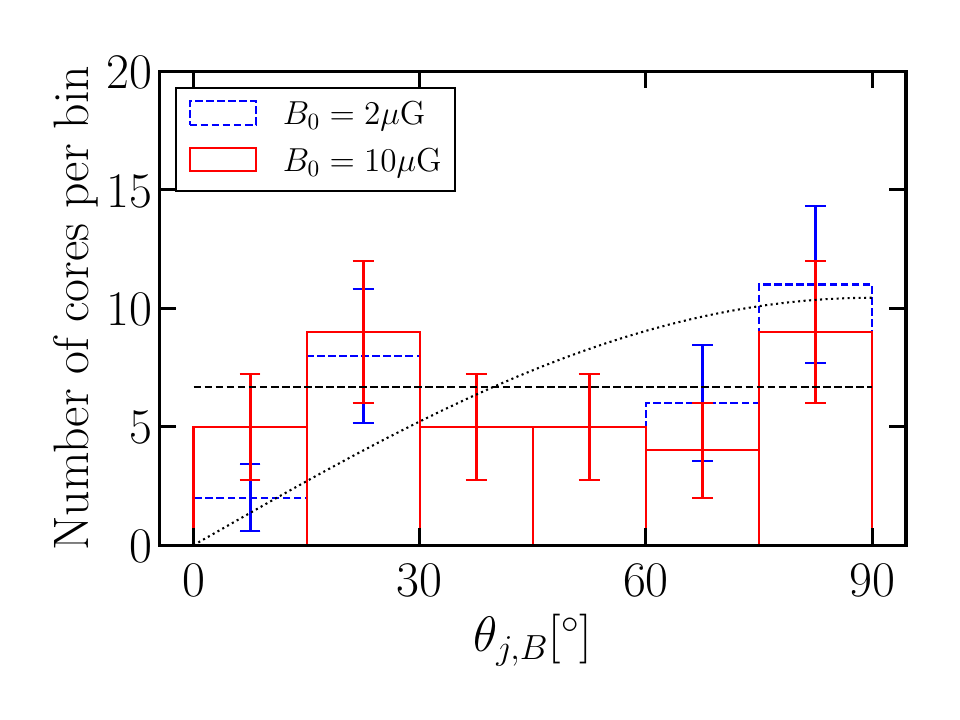}{0.50\textwidth}{(c) $3.0 \ {\rm M}_{\odot}$}
          }
\caption{Histogram of the angle between the angular momentum vector and the magnetic field of the cores at the final state with (a) $0.1 \ {\rm M}_{\odot}$, (b) $1.0 \ {\rm M}_{\odot}$, and (c) $3.0 \ {\rm M}_{\odot}$. The blue dashed and red solid lines represent the case of weak and strong magnetic field cases, respectively. The black dotted line corresponds to the random distribution in the three-dimensional space taking into account the effect of the solid angle. The black dashed line represents the random distribution in the 2D plane perpendicular to the filament axis. }
\label{fig:thetajb_hist}
\end{figure*}

In section \ref{subsec:anisoAMtransfer}, we show that the anisotropy of the angular momentum transfer due to the magnetic field can be found in the mass scale $M_{\rm ana} \lesssim 1.0 \ {\rm M}_{\odot}$. Since the angular momentum is efficiently removed in the perpendicular case compared to the parallel case \citep{Mouschovias1985}, it is expected that the core rotation axis tends to be parallel to the local magnetic field direction at the final state.  To confirm whether this is true or not, in this subsection, we investigate the distribution and evolution of the 3D angle between the angular momentum and the magnetic field in more detail.\par
First, in Figure \ref{fig:thetajfil_hist}, we show the distribution of the angle between the angular momentum of the core and the filament axis at the final state of our simulations. Here, the $z$-axis (the initial filament axis) is used as the reference axis for measuring the rotation axis. Figure \ref{fig:thetajfil_hist} shows that the rotation directions of the cores are mostly perpendicular to the filament axes irrespective of the initial strength of the magnetic field. The reason for this is the same in the case of non-magnetized filaments shown in Paper\Rnum{1}: this is because the cores are elongated along the $z$-axis (filament longitudinal direction) at the initial state (Figure \ref{fig:lagtrace}). In the case of $M_{\rm ana}=0.1\ {\rm M}_{\odot}$, the tendency of the perpendicular rotation is weak compared with the case of $M_{\rm ana}=3.0\ {\rm M}_{\odot}$ because the effect of the filament geometry is less effective for smaller mass scale. \par
Figure \ref{fig:thetajb_hist} displays the distribution of the angle between the angular momentum and the local magnetic field, $\theta_{j, B}$ at the final state. The black dotted line in Figure \ref{fig:thetajb_hist} represents the random distribution in the three-dimensional space. The distribution increases toward larger $\theta$ because of the solid angle. We refer to the black dotted line as the 3D random distribution. On the other hand, if both rotation and magnetic field vectors are random but confined in the 2D plane perpendicular to the filament axis, the distribution is flat. This flat distribution is described as the black dashed line in Figure \ref{fig:thetajb_hist}. Since, at the initial state, the magnetic field lines are parallel to the $x$-axis and the rotation axis of the core is randomly oriented in the $x$-$y$ plane due to the elongation of the cores along the filament longitudinal axis (Figure \ref{fig:lagtrace}), the angle between the rotation axis and the magnetic field line follows the flat distribution at the initial state. We note that the direction of the magnetic field within the core also changes with time due to the turbulent velocity field and the gravitational fragmentation of the filament. Figure \ref{fig:thetajb_hist} indicates that the distribution of $\theta_{j, B}$ is flatter than the 3D random distribution. This means that the magnetic field lines tend to be confined in the $x$-$y$ plane compared with the 3D random distribution. Even at $M_{\rm ana} = 0.1 \ {\rm M}_{\odot}$, 11 cores have $\theta_{j ,B}>60^{\circ}$. As shown in Figure \ref{fig:thetajfil_hist}, most cores rotate perpendicular to the filament axis. In the strong magnetic field case, the magnetic field direction of half of the cores in the sample remains perpendicular even just before the first core formation (Section \ref{subsec:synobs}). Therefore, in the strong magnetic field case, both rotation vector and the magnetic field direction of half of the cores are roughly on the $x$-$y$ plane which is perpendicular to the filament axis. On the other hand, in the weak magnetic field case, since the magnetic field direction significantly changes during the core formation phase (Section \ref{subsec:synobs}), the magnetic field lines are not confined in the $x$-$y$ plane. In the weak magnetic field case, although the rotation axis of the cores tends to be perpendicular to the filament axis, the magnetic field lines show a parallel component along the filament axis at the core scale. This means that the distribution of $\theta_{j, B}$ tends to have a 3D random distribution in the three dimension rather than on the $x$-$y$ plane compared with the strong magnetic field case. \par
These results of strong magnetic field case indicate that the rotation axis of the cores is not always aligned with the local magnetic field direction. To reveal what causes this diversity, we focus on the two cores in the perpendicular sample that have the angle between the angular momentum and the local magnetic field larger than $60^{\circ}$ at $\rho_{\rm c}=10^{-18} \ {\rm g \ cm^{-3}}$. At the final state, one is aligned to the magnetic field, the other has large inclination between the rotation axis and the magnetic field direction even though both cores are in the perpendicular sample. The former and latter cores are referred to as aligned core and misaligned core, respectively. \par
\begin{figure*}
\gridline{\fig{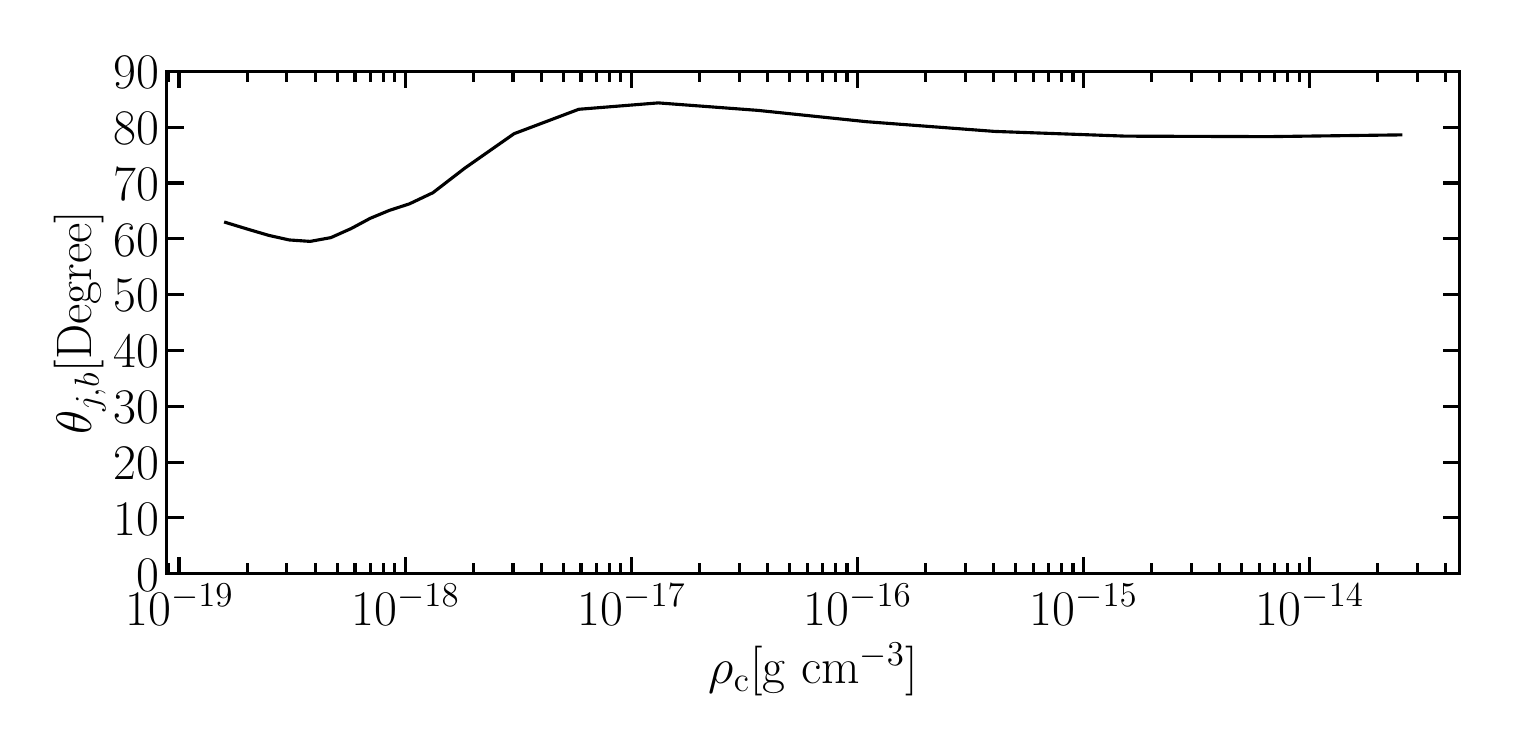}{0.49\textwidth}{(a)}
	\fig{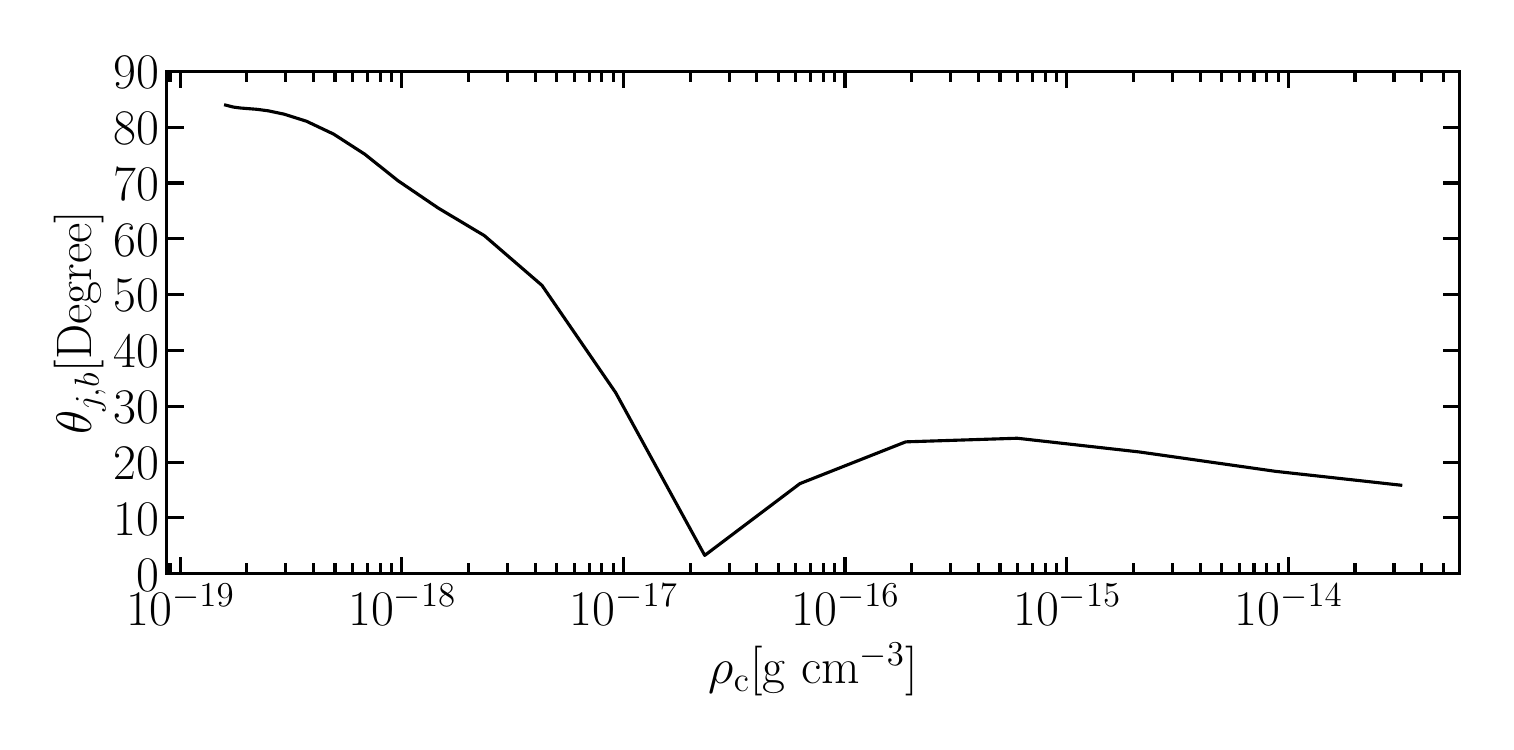}{0.49\textwidth}{(b)}
          }
          \gridline{\fig{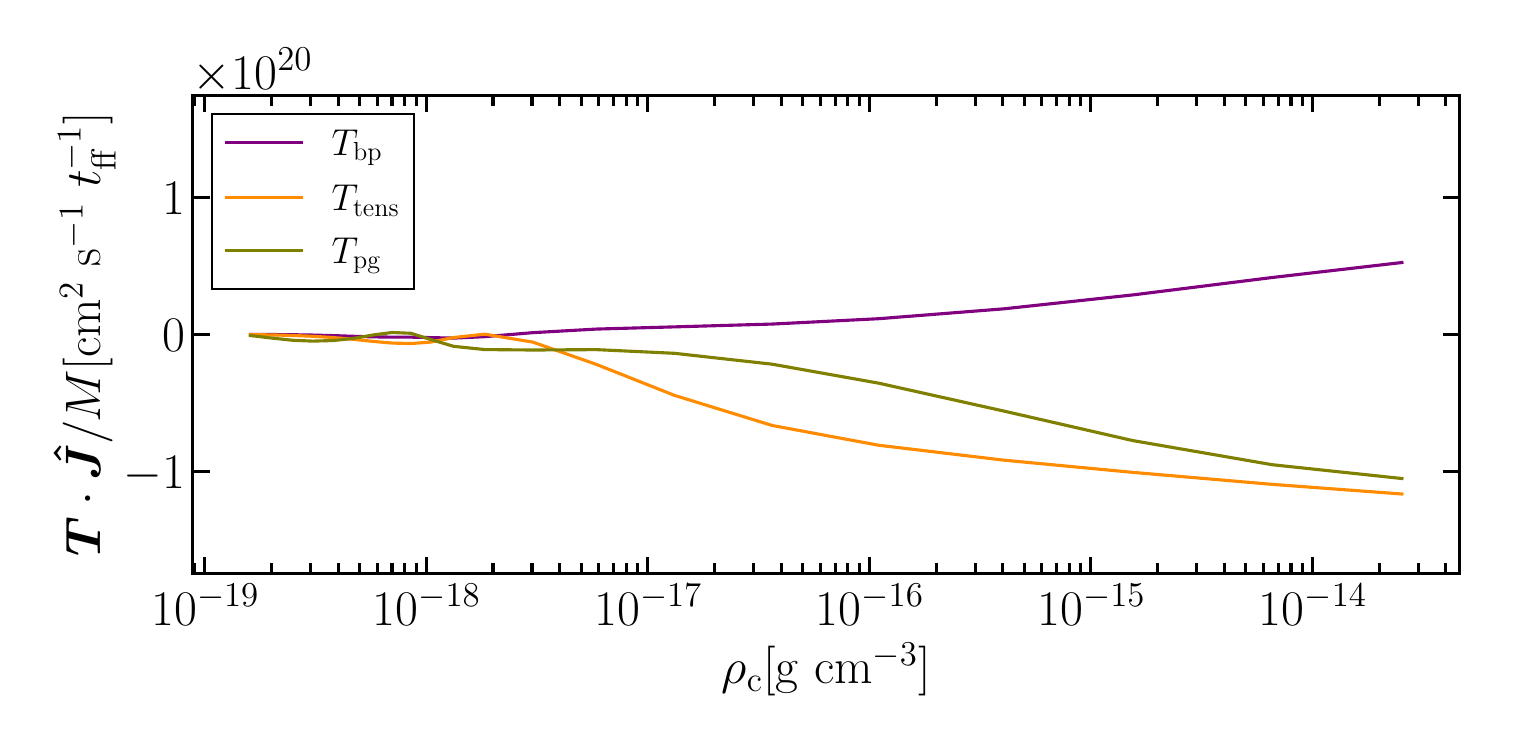}{0.49\textwidth}{(c)}
	\fig{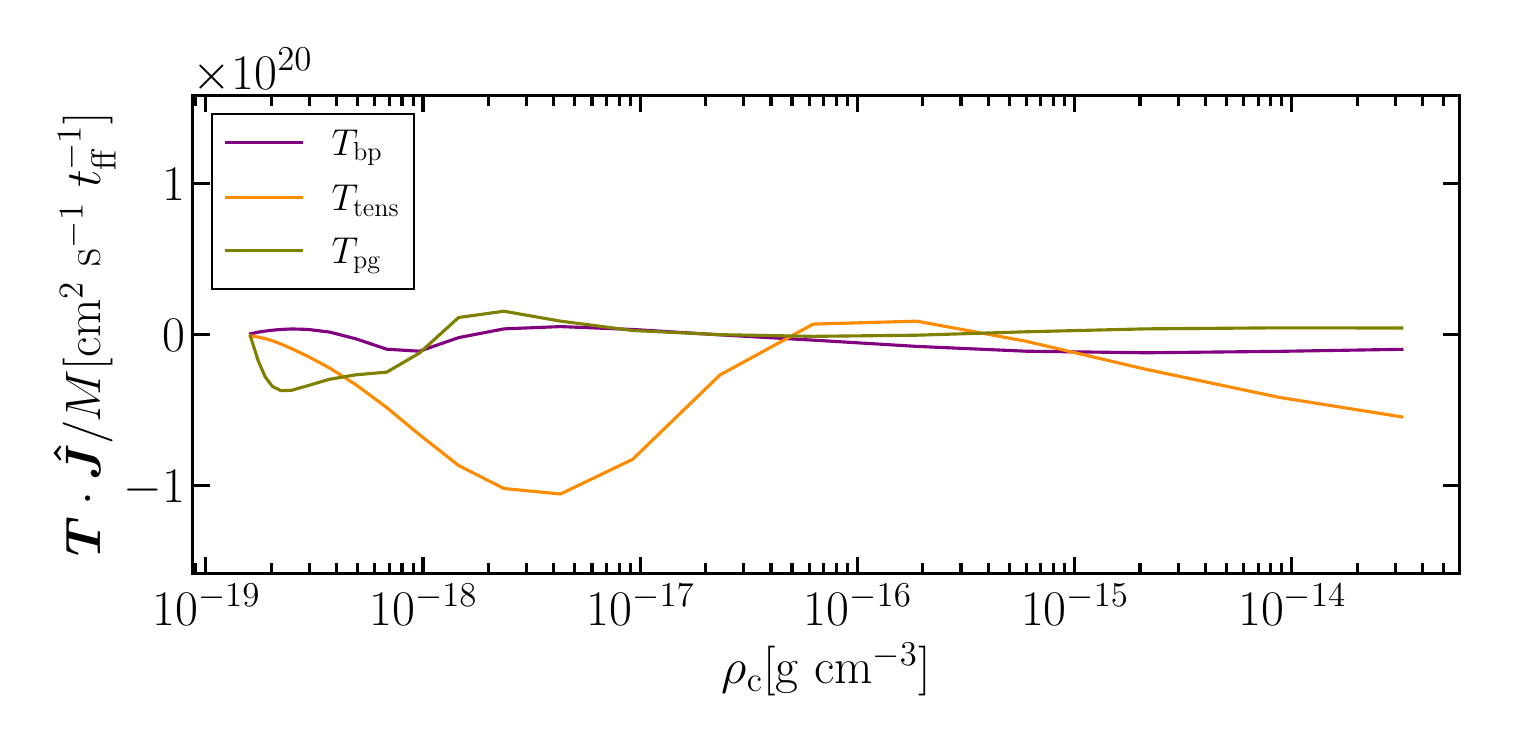}{0.49\textwidth}{(d)}
          }
\gridline{\fig{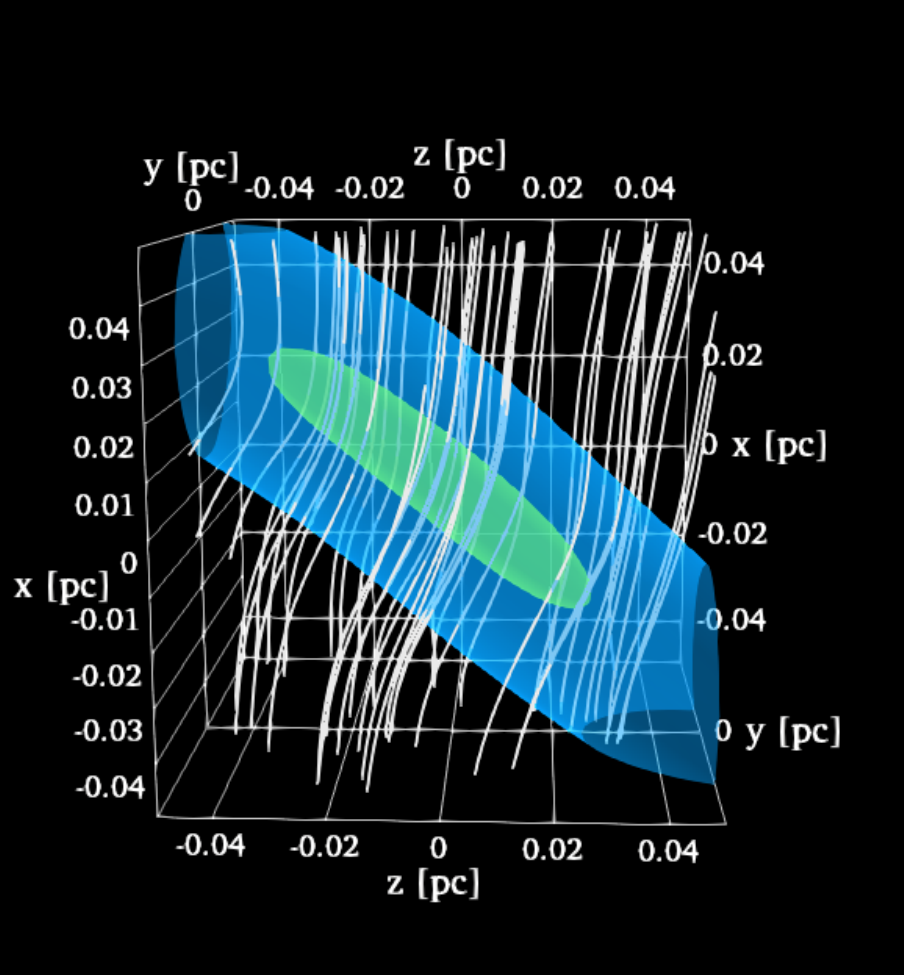}{0.49\textwidth}{(e)}
          \fig{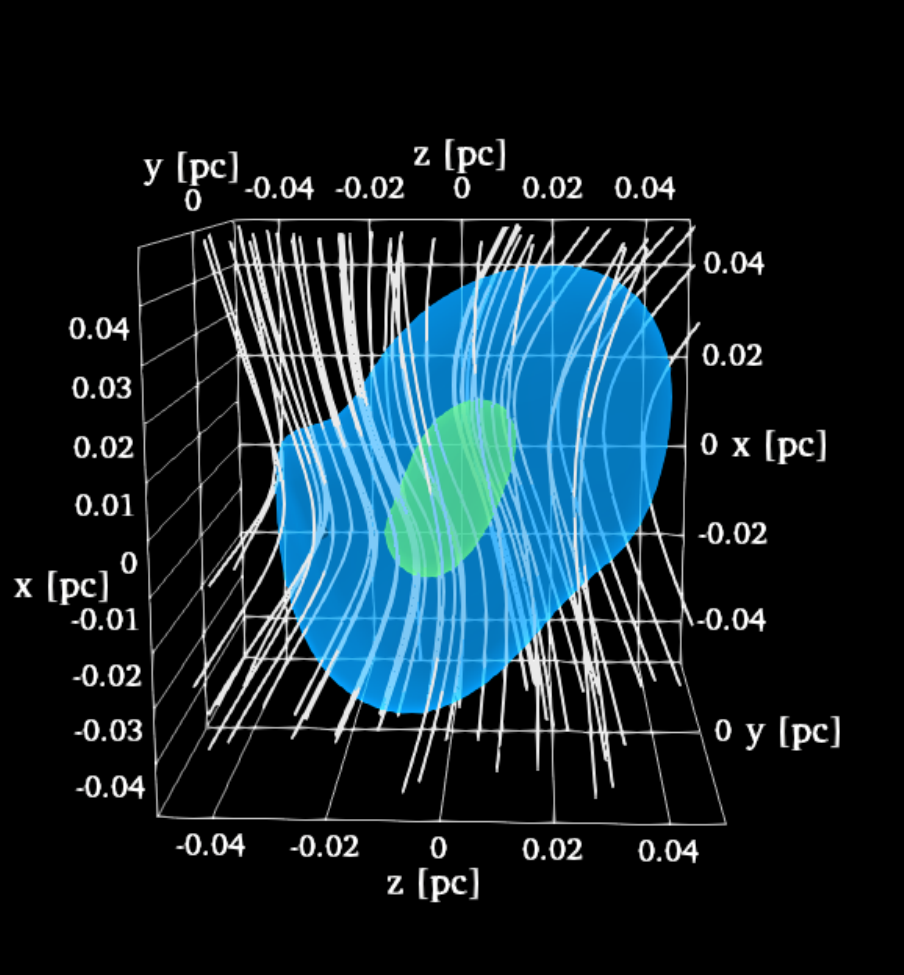}{0.49\textwidth}{(f)}
          }

\caption{Comparison of the aligned (first column - a, c, e) and misaligned (second column - b, d, f) cases. The first row is the evolution of the angle between the angular momentum and the local magnetic field direction at $M_{\rm ana} = 0.1 \ {\rm M}_{\odot}$. The second row shows the evolution of the torques exerted on the cores at $M_{\rm ana} = 0.1 \ {\rm M}_{\odot}$. The vertical axis represents the inner product between the torque and the unit vector of the angular momentum. The third row is the density (color) and magnetic field (white lines) structures in the cores. The blue and green isosurfaces represent the isodensity surfaces, $\rho=1.7 \times 10^{-19} \ {\rm g \ cm^{-3}}$ ($n=4.4 \times 10^4  \ {\rm cm^{-3} }$) and $\rho=9 \times 10^{-19} \ {\rm g \ cm^{-3}}$ ($n=2.3 \times 10^5 \ {\rm cm^{-3} }$), respectively. \label{fig:comp_align_ornot} }
\end{figure*}

The time evolution of $\theta_{j, B}$ at $M_{\rm ana} = 0.1 \ {\rm M}_{\odot}$ is shown in the top panels of Figure \ref{fig:comp_align_ornot}. The rotation axis of the misaligned core remains perpendicular to the local magnetic field direction. However, the aligned core experiences a strong alignment when $10^{-18} \ {\rm g \ cm^{-3}} < \rho_{\rm c} < 10^{-17} \ {\rm g \ cm^{-3}}$. The middle panels of Figure \ref{fig:comp_align_ornot} displays the evolution of the torques exerted on each core at $M_{\rm ana} = 0.1 \ {\rm M}_{\odot}$. As shown in panel (c) of Figure \ref{fig:comp_align_ornot}, the magnetic tension torque exerted on the misaligned core monotonically increases with time. On the other hand, in the case of aligned core, the magnetic tension torque rapidly increases until the central density reaches $\rho_{\rm c} \simeq 10^{-17} \ {\rm g \ cm^{-3}}$. Since the timescale becomes shorter as the gravitational collapse proceeds, the torque in the early evolutionary stage transfers the angular momentum more efficiently compared to that at the later evolutionary stage. Therefore, the strong magnetic tension torque acting on the aligned core during $ 10^{-18} \ {\rm g \ cm^{-3}} < \rho_{\rm c} < 10^{-17} \ {\rm g \ cm^{-3}}$ leads to strong alignment between the rotation axis and the local magnetic field direction shown in panel (b) of Figure \ref{fig:comp_align_ornot}. The magnetic filed structure is shown in the bottom panels of Figure \ref{fig:comp_align_ornot}. The magnetic field lines are relatively straight in the case of the misaligned core while the mean direction is inclined with the initial magnetic field direction ($x$-axis) due to the large eddy motion. On the other hand, the magnetic filed lines are strongly pinched in the case of the aligned core. Our results indicate that the pinched magnetic filed configuration causes the strong magnetic braking. This is consistent with the analytical discussion in \cite{Mouschovias1985}. The pinched geometry has a longer lever arm compared to the case of the misaligned core. These results imply that the cores formed in the filament with the same line mass and the same initial magnetic filed have a diversity of the magnetic field structure just before the first core formation due to the difference of the initial seed of the turbulence even in the case of transonic initial turbulence.

\section{Discussion} \label{sec:discussion}

\subsection{Direction of the Magnetic Field \label{subsec:synobs}}


\begin{figure*}
\gridline{
	\fig{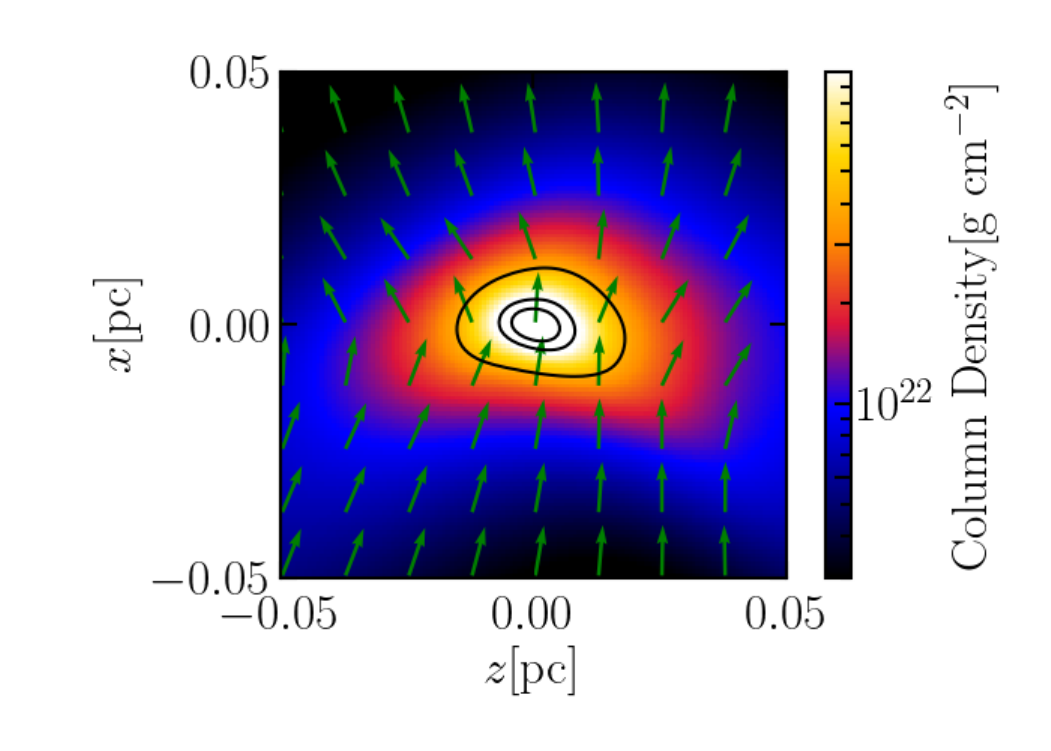}{0.33\textwidth}{(a)}
        \fig{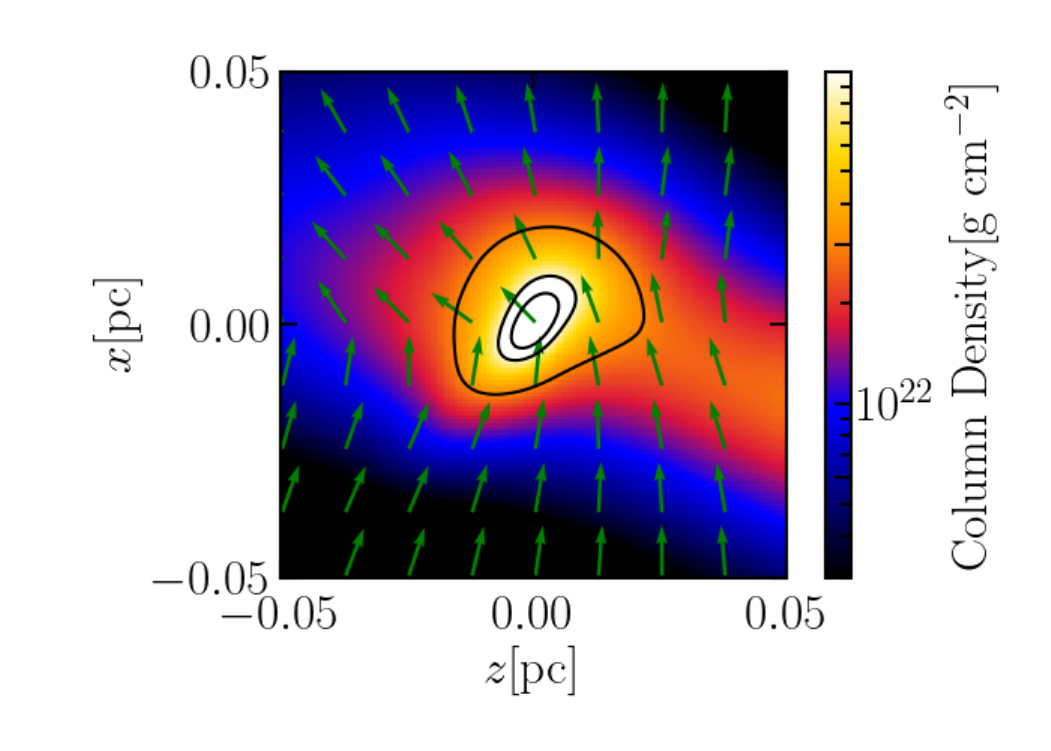}{0.33\textwidth}{(b)}
         \fig{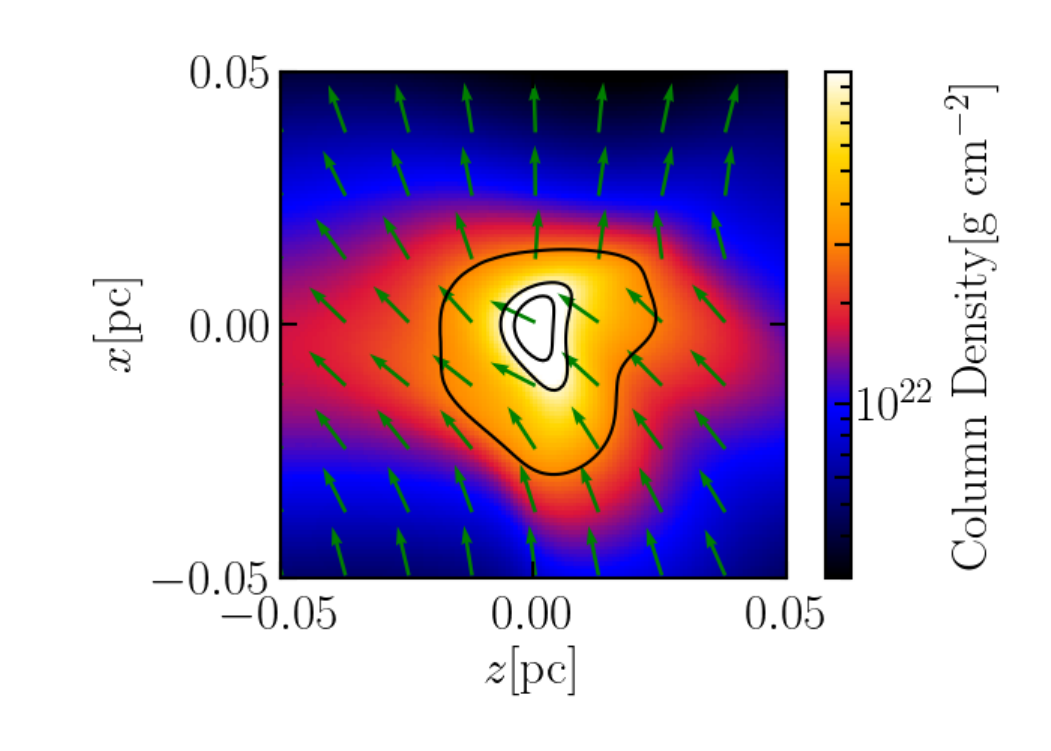}{0.33\textwidth}{(c)}
          }
\caption{Synthetic observations of magnetic fields in the cores. The color and green arrow represent the column density and the magnetic field direction derived by integrating the density and magnetic field (using Equation \ref{eq:bobscal}) along the $y$ direction. The contours correspond to 50\%, 30\%, 10\% of the peak column density.  \label{fig:magsynth} }

\end{figure*}

\begin{figure}
\epsscale{0.7}
\plotone{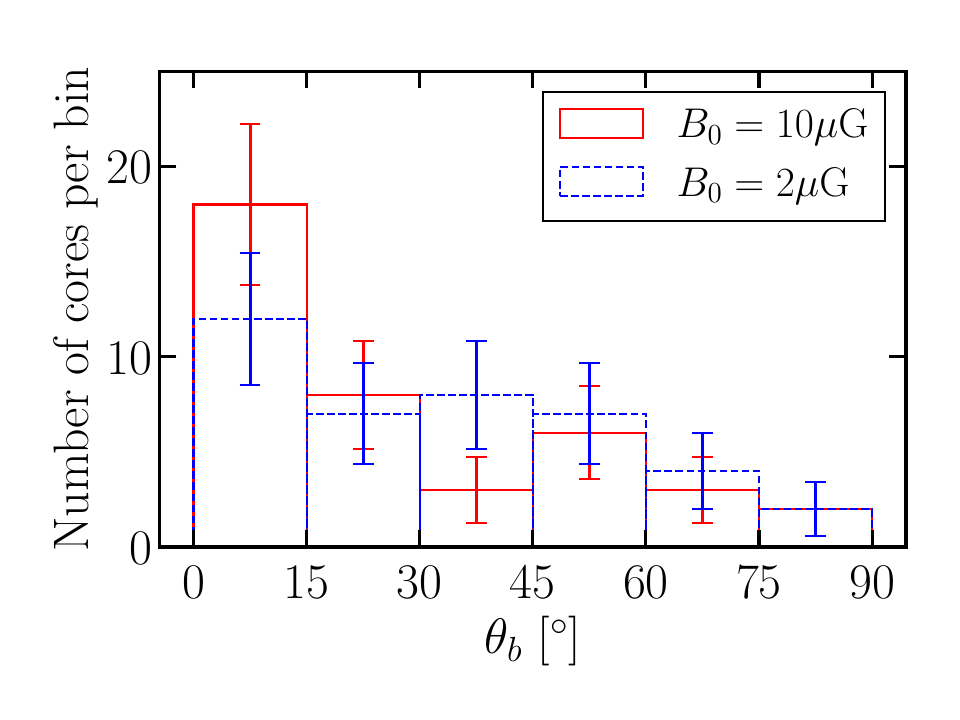}
\caption{Histogram of the angle between the direction of the magnetic field measured in the synthetic observation map and the $x$-axis which corresponds to the initial direction of the magnetic field. The direction of the magnetic field is measured when the maximum density of the core reaches $  \rho_{\rm c} = 10^{-16} \ {\rm g \ cm^{-3}} $ \label{fig:costhetab_synth} }
\end{figure}

At the initial state of our simulations, the filaments are perpendicularly threaded by the magnetic field. However, the turbulent motion of the gas can deform the magnetic field lines and hourglass magnetic field structures are created by the longitudinal contraction motions due to gravity. Observations show that the magnetic field directions towards the cores formed along the same filament show a variety of orientation compared to the filament axis \citep{Eswaraiah2021}. Therefore, it is important to study whether our results of simulations can explain the diversity of the configuration of magnetic field lines observed in molecular cloud cores along a given filament or not. To make synthetic observational-like maps of the plane of the sky component of the magnetic field as traced by observations of the dust polarized emission, we integrate the magnetic field along the $y$-direction weighted by density as follows:
\begin{eqnarray}
\boldsymbol{B}_{\rm obs}= \frac{\int \rho \boldsymbol{B} dy}{\int \rho dy}.
\label{eq:bobscal}
\end{eqnarray}
Figure \ref{fig:magsynth} displays the synthetic observations of our simulations towards a few examples of cores. In Figure \ref{fig:magsynth} (a), one can clearly see the hourglass structure formed by the filament fragmentation and the contraction of the core due to gravity. Figure \ref{fig:magsynth} (b) and (c) show that, although the magnetic field lines are relatively straight in the outer low density region, the magnetic field lines in the central high density region of the cores is bent and not aligned with the initial magnetic field direction due to the core rotation motion inherited from the initial turbulent velocity field. Figure \ref{fig:costhetab_synth} displays the histogram of the angle between the initial magnetic filed direction and the local magnetic field direction in the high density region enclosed by the contour of $50 \%$ of the peak column density. The histogram shown in Figure \ref{fig:costhetab_synth} suggests that the magnetic field direction in the cores has a diversity, and this result is compatible with the observations which reports the variety of the magnetic field directions inside the core \citep{Eswaraiah2021}. 
Figure \ref{fig:costhetab_synth} also suggests that the histogram of weak magnetic field case shows a flat distribution compared with that of strong magnetic field case. On the other hand, the local magnetic field direction of half of the cores in the strong magnetic field case is aligned with the initial magnetic field direction ($ 0^{\circ} < \theta_{\rm b}<15^{\circ}$). Our result indicates that the statistical measurement of the angle between the filament axis and the local magnetic field direction will give us a hint to determine the strength of the magnetic field.

\subsection{Effect of Selective Accretion \label{subsec:sele}}

\begin{figure*}
\gridline{\fig{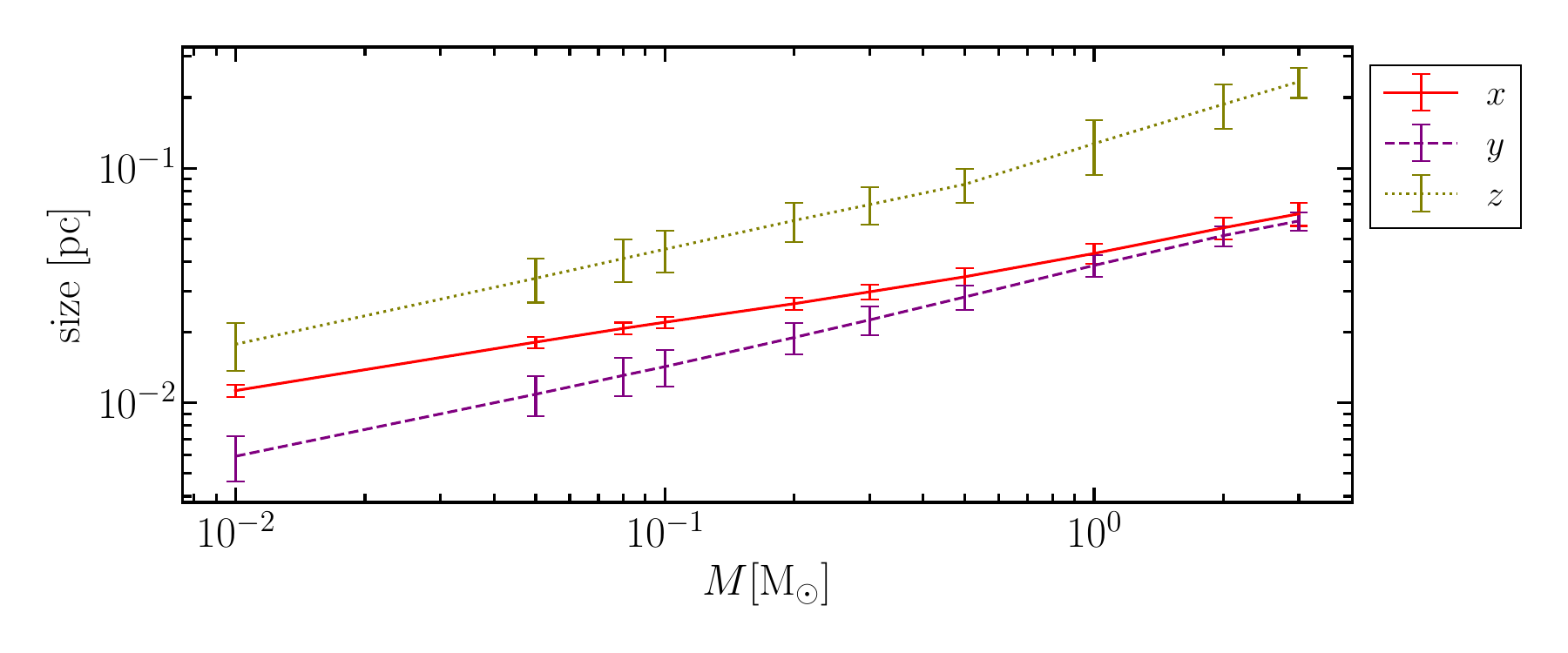}{0.45\textwidth}{(a)}
          \fig{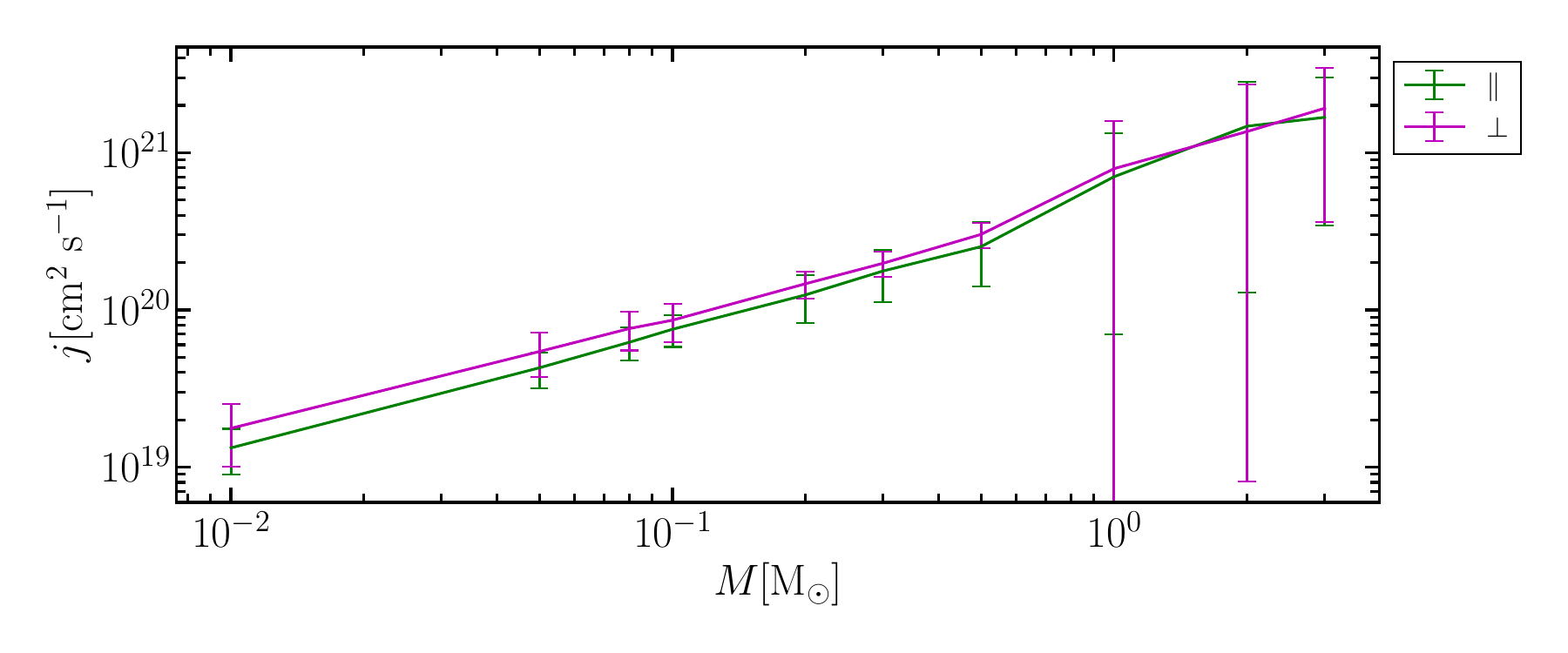}{0.45\textwidth}{(b)}
          }
\caption{Initial shape (a) and specific angular momentum (b) of the cores in the case of strong magnetic field. The vertical axis of panel (a) and (b) are the size and the specific angular momentum of cores at the initial state, respectively. The horizontal axis represents the core mass. The red solid, purple dashed, and olive dotted lines in the panel (a) are the core size along the $x$, $y$, and $z$-axes at the initial state, respectively. The green and magenta solid lines in the panel (b) are the evolution of the specific angular momentum of the cores of parallel and perpendicular samples at the initial state, respectively. }
\label{fig:selection}
\end{figure*}

In Section \ref{subsec:anisoAMtransfer}, we show that the distribution of the angle between the local magnetic field and the rotation axis is random within the $x$-$y$ plane in the strong magnetic field case. However, if the magnetic field is strong enough, the gas falls onto the core along the initial magnetic field direction ($x$-axis).  This means that the strong magnetic field could cause the anisotropy of the angular momentum of the core even at the initial state of our simulations. To investigate the effect of selective infall, we derive the initial shape of the cores as shown in Figure \ref{fig:selection} (a). The size of the cores along each direction is defined as, for example $x_{\rm max}-x_{\rm min}$, where $x_{\rm max}$ and $x_{\rm min}$ are the maximum and minimum value of $x$ of the SPH particles contained in the core, respectively. The size of the $y$ and $z$ directions are defined using the same method. Figure \ref{fig:selection} (a) suggests that the effect of the magnetic field on the shape of the core can be clearly seen in the inner region. This means that the selective infall due to the magnetic field occurs in the case of strong magnetic field. The anisotropy of the shape can be seen only in the inner region since the contraction due to the gravity occurs only in the inner region. The outer region which traces the filament is mainly supported by the thermal pressure. \cite{Tsukamoto2018} showed that the selective accretion affects the evolution of the angular momentum of the central high density region. If the angle between the rotation axis and the magnetic field lines is parallel, the fluid elements with smaller angular momentum fall onto the central region. On the other hand, if the misalignment angle is large, the gas elements with larger angular momentum selectively accrete onto the central region. Figure \ref{fig:selection} (b) displays the initial specific angular momentum at each mass scale. The green and magenta solid lines represent the specific angular momentum of parallel and perpendicular samples at the initial state, respectively. The difference between both samples is relatively small in all mass scales. This is because the cores are elongated along the $z$-axis even in the small mass scale. The elongation along the $z$-axis hides the effect of the selective accretion guided by the strong magnetic field due to the flux-fleezing in our ideal MHD simulations. The evolution of the angular momentum due to selective accretion should be tested in the future with non-ideal MHD simulations.

\subsection{Analytical Estimate of the Critical Mass of Magnetic Braking  \label{subsec:anatraindis}}

\begin{figure}
\epsscale{0.85}
\plotone{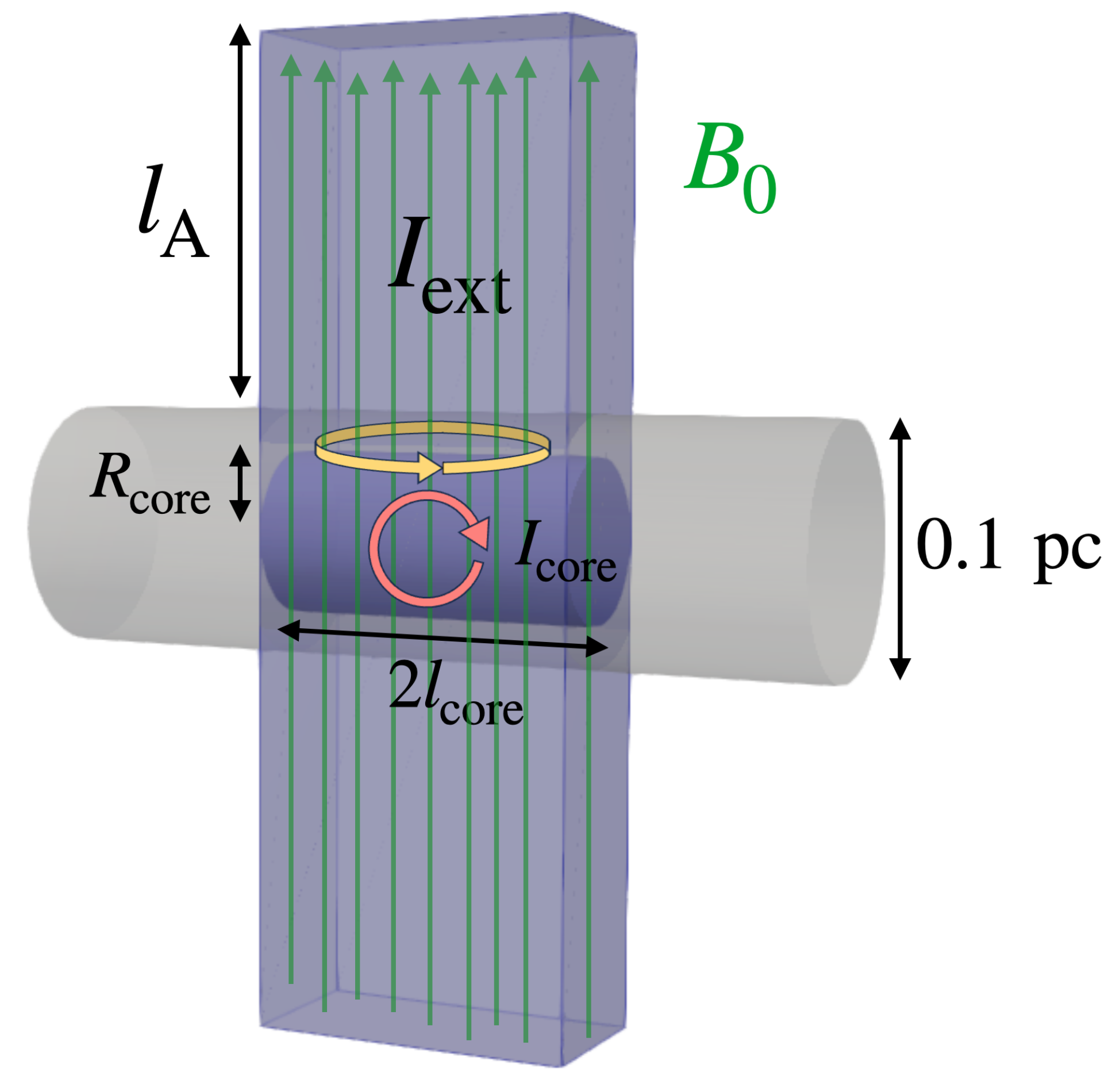}
\caption{Schematic figure for the analytical model. The rotation motions of the perpendicular and parallel configurations are described as the red and yellow arrows, respectively. \label{fig:Analytical_schematic} }
\end{figure}

\begin{figure*}
\gridline{\fig{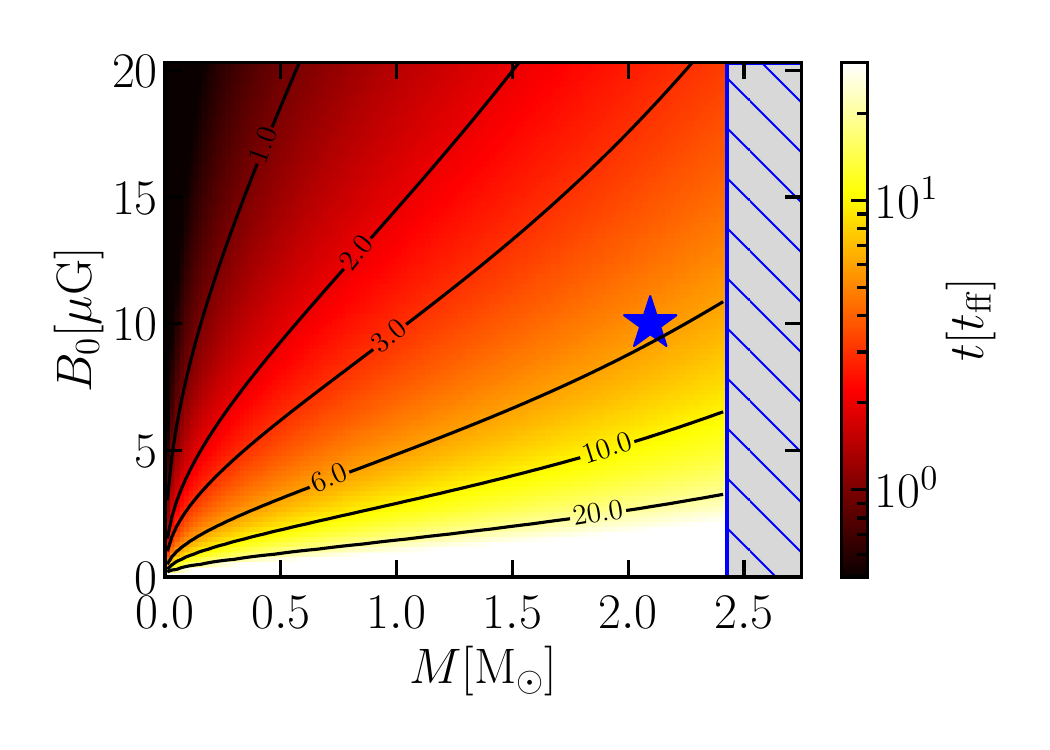}{0.49\textwidth}{(a)}
          \fig{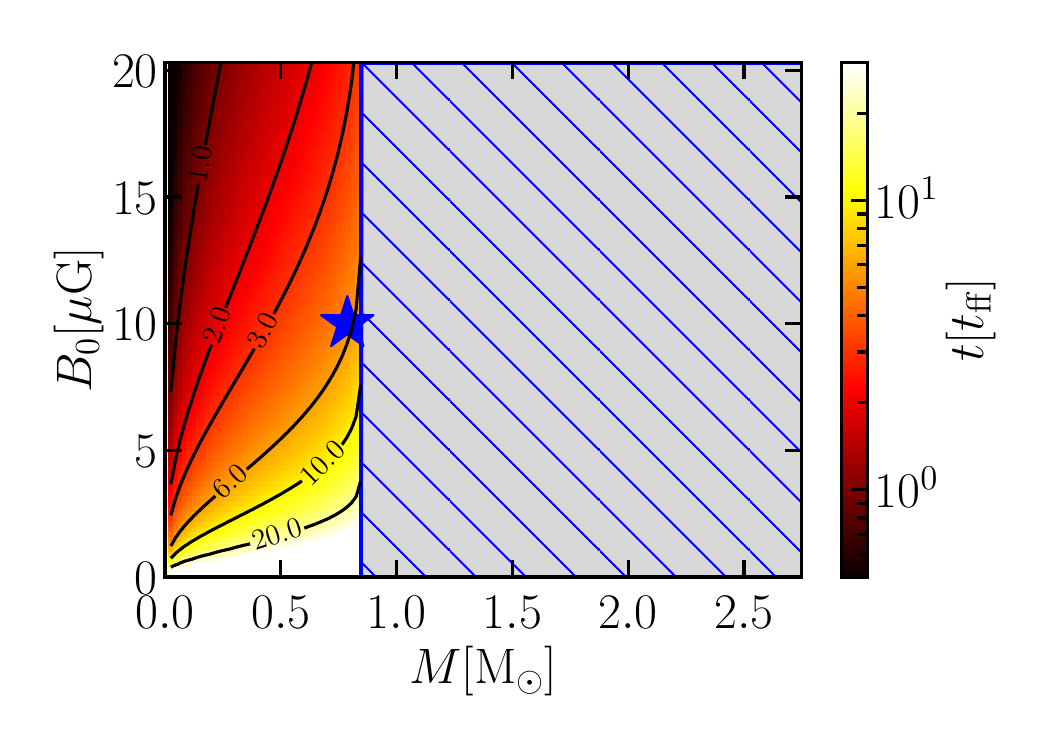}{0.49\textwidth}{(b)}
          }
\caption{Timescale of magnetic braking derived from our semi-analytical model. The left and right panels represent the timescale of the magnetic braking in the case of perpendicular and parallel configurations, respectively. The vertical and horizontal axes represent the strength of the magnetic field and the core mass of the model, respectively. In the shaded region, the magnetic braking is not efficient even when the Alfv\'en wave reaches infinity. The blue stars represent the critical mass in the case of the strong magnetic field when we adopt the elapsed time of our simulation, $t_{\rm fin} = 5.54\ t_{\rm ff}$, as the propagation time of the Alfv\'en wave. }
\label{fig:ana_timescale}
\end{figure*}

In Section \ref{subsec:AMtransfer}, we define $M_{\rm crit}$ below which the angular momentum transfer by the magnetic field is not important, and our results show $1.0 \ {\rm M}_{\odot} \lesssim M_{\rm crit} \lesssim 3.0  \ {\rm M}_{\odot}$. To study the origin of $M_{\rm crit}$, we calculate the timescales of magnetic braking based on the concept of the analytical model shown in \cite{Mouschovias1979,Mouschovias1980} and \cite{Mouschovias1985}. The timescale of magnetic braking of the core is estimated as the time needed for the Alfv\'en wave to sweep a region that has the same inertia as the core. To calculate this timescale, we consider a simple model as shown in Figure \ref{fig:Analytical_schematic}. In this model, the core is a concentric cylindrical region in the filament whose length and radius are written as $2l_{\rm core}$ and $R_{\rm core}$, respectively. The magnetic field is uniform and perpendicular to the filament axis. The filament axis and magnetic field direction are along the $z$- and $x$- axes, respectively. We adopt the same hydrostatic density profile of the initial condition of our simulations (Equation \ref{eq:inidenpro}). The Alfv\'en wave propagates from the boundary between the core and the surrounding medium along the magnetic field described as the light blue region in Figure \ref{fig:Analytical_schematic}. $l_{\rm A}(t)$ represents the distance propagated by the Alfv\'en wave during the time $t$. In the model, we consider two configurations. First, the perpendicular configuration where the rotation direction of the core is perpendicular to the magnetic field direction and to the filament axis. Second, the parallel configuration where the rotation vector is parallel to the magnetic field and perpendicular to the filament axis. In Figure \ref{fig:Analytical_schematic}, the rotation motions of the perpendicular and parallel configurations are described as the red and yellow arrows, respectively. The equations relative to the moment of inertia of the core and the external medium in our model are shown in Appendix A. Figure \ref{fig:ana_timescale} displays the timescale of magnetic braking derived from our semi-analytical model. Since we adopt the density profile of hydrostatic equilibrium, the moment of inertia has finite value in both perpendicular and parallel cases even when $l_{\rm A} \rightarrow \infty$.
Moreover, the Alfv\'en wave reaches infinity in a finite time which is described as follows:
\begin{eqnarray}
\label{eq:traveltime}
\nonumber t_{\rm A} & = & \displaystyle \int^{\infty}_{0}\frac{ds}{v_{A}(s)} \\  
& = & \displaystyle \frac{\pi H_0}{2v_{\rm A}(0)}, 
\end{eqnarray}
where $v_{\rm A} (s) \equiv B_0/\sqrt{4\pi \rho(s)}$ is the Alfv\'en wave. The region in which the magnetic braking is ineffective is described by the shaded region in Figure \ref{fig:ana_timescale}. The critical mass for the cores in the parallel case is smaller than that of the perpendicular cores since the moment of inertia in the perpendicular case is larger than that of the parallel case. This is consistent with the results of \cite{Mouschovias1985}. The average elapsed time from initial to final state of our simulation in the strong magnetic filed case is $t_{\rm fin} = 5.54\ t_{\rm ff}$, which includes the timescale of the filament fragmentation and that of the collapse of the core. Using the elapsed time $t_{\rm fin}$ as the timescale of the magnetic braking and adopting the initial strength of the magnetic field of $10 \ {\rm \mu G}$, we can determine the critical mass scale from the Figure \ref{fig:ana_timescale}. The critical masses in the case of perpendicular and parallel configuration are $M_{\rm crit} = 2.10$ and $0.79 \ {\rm M}_{\odot}$\footnote{$t_{\rm fin}$ depends on the initial seed of turbulence. In our simulations, the standard deviation of $t_{\rm fin}$ is $1.25\ t_{\rm ff}$. If we adopt $t_{\rm fin} = 4.29\ t_{\rm ff}$, the resultant critical mass in the case of perpendicular and parallel configuration are $1.62 \ {\rm M}_{\odot}$ and $0.66 \ {\rm M}_{\odot}$, respectively. On the other hand, if we use $t_{\rm fin} = 6.79\ t_{\rm ff}$, the critical mass in the case of perpendicular and parallel configuration are $2.43 \ {\rm M}_{\odot}$ and $0.84 \ {\rm M}_{\odot}$, respectively. These results suggest that the choice of $t_{\rm ff}$ does not make a significant impact on our conclusion in which the magnetic braking is not important $M \gtrsim 3.0 \ {\rm M}_{\odot}$.}. These critical masses are plotted as the blue stars in Figure \ref{fig:ana_timescale}. These results are consistent with the conclusions from our simulations in which the magnetic braking is not important for $M \gtrsim 3.0  \ {\rm M}_{\odot}$. In summary, the magnetic filed does not make a significant impact on the angular momentum transfer at the core scale ($\sim 0.1$ pc).

\subsection{Implication for Diversity of Circumstellar Disk\label{subsec:diversity}}

\begin{figure}
\epsscale{0.9}
\plotone{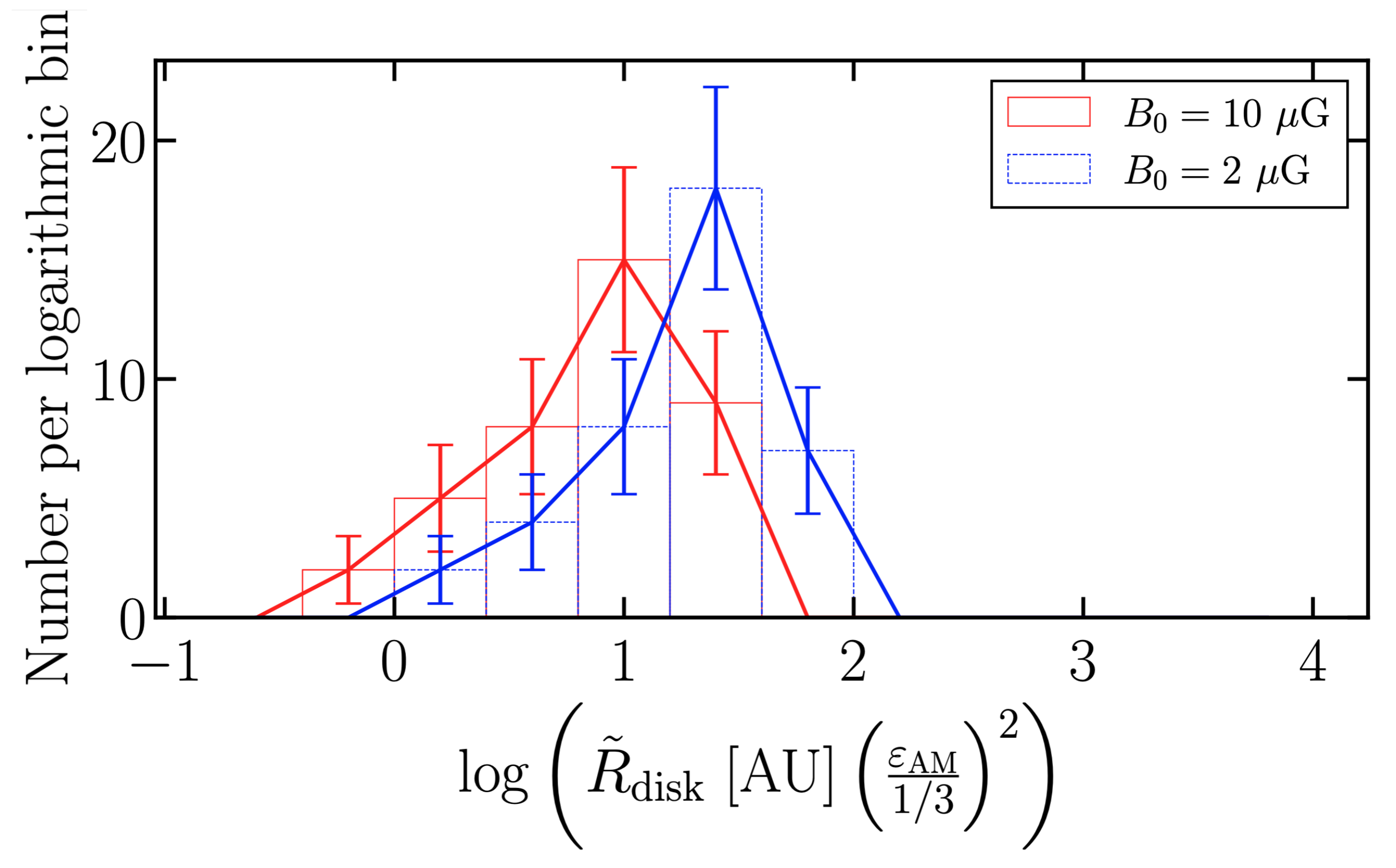}
\caption{Histogram of the disk radii predicted by our simulations. The red solid and blue solid lines are the histograms of disk radius in the case of $B_0=10 \ {\rm \mu G}$ and $B_0=2 \ {\rm \mu G}$ estimated from Equation \ref{eq:rdisk}. \label{fig:disksize} }
\end{figure}

Recent observational results using thermal dust emission at (sub-)millimeter wavelengths suggest that the circumstellar disks around young stellar objects,  such as Class 0 and Class\Rnum{1} objects, have a large diversity in size and mass of protoplanetary disks \cite[e.g.,][]{Cox2017,Maury2019,Tobin2020}. For a recent detailed review see \cite{Tsukamoto2022}. \cite{Tobin2020} show that the mean disk radii of Class 0, Class\Rnum{1}, and flat spectrum protostars measured from dust continuum emission in the Orion molecular clouds are 44.9, 14.9, and 28.5 {\rm AU}, respectively. They also reported that the dispersion of the distribution of disk radii of Class 0, Class\Rnum{1}, and flat spectrum protostars are $\sigma\left(\log _{10}\left(R_{\mathrm{disk}}/(1 \ \mathrm{AU})\right)\right) = 0.38, 0.42, \text{and} \ 0.38$, respectively. To investigate an origin of this diversity, \cite{Kuffmeier2017} and \cite{Bate2018} performed simulations starting from the molecular cloud scale with initial turbulent velocity field. Although the distribution of disk radius shown in \cite{Bate2018} seems to be consistent with the observations, the simulation of \cite{Bate2018} does not include the effect of magnetic field. \cite{Kuffmeier2017} performed zoom-in simulations with the maximum resolution reaching 2 AU. Although the resolution of their simulations is relatively high, the sample size of the disk is not large enough. \par
We define the ``equivalent angular momentum radius", $\tilde{R}_{\rm disk}$, by the following equation:
\begin{eqnarray}
M_{\rm disk} \sqrt{ G M_{\rm star} \tilde{R}_{\rm disk} } = \varepsilon_{\rm AM} J_{\rm tot},
\label{eq:rdisk}
\end{eqnarray}
where $M_{\rm star}$ and $M_{\rm disk}$ are the mass of the central star and disk, respectively. $\varepsilon_{\rm AM}$ is the parameter to control the angular momentum transfer from the first core formation epoch to the disk formation epoch. $\varepsilon_{\rm AM}$ is defined as the ratio of the specific angular momentum of the core at the final state of our simulations to that of the disk at the disk formation epoch. At the final state of our simulations, which is just before the first core formation, we measure the total angular momentum, $J_{\rm tot}$, of the spherical region with the enclosed mass of $1.0 \ {\rm M}_{\odot}$. Then, we calculate $\tilde{R}_{\rm disk}$ using Equation \ref{eq:rdisk}. To calculate $\tilde{R}_{\rm disk}$ from Equation \ref{eq:rdisk}, $M_{\rm star}$ and $M_{\rm disk}$ should be determined. Roughly speaking, one third of the core mass is ejected by the outflow and another one third of the mass in the outer region of the core tends to escape from the core because it becomes gravitationally unbound by the loss of the system mass due to the outflow \cite[e.g.,][]{Machida2009,Machida2012,Machida2013}. Then, we define the disk-to-star mass ratio, $\varepsilon_{\rm disk}=M_{\rm disk}/M_{\rm star}$. Here, we adopt $\varepsilon_{\rm disk}=0.1$ \cite[e.g.,][]{Gammie2001}. The correct value of $\varepsilon_{\rm AM} (<1) $ may depend on the property of each core such as the magnetic field strength, magnetic field orientation, and gravitational energy. Here we simply adopt $\varepsilon_{\rm AM} = 1/3 $ as a typical value found in protostellar collapse calculations \cite[e.g.,][]{Tomisaka2000, Tsukamoto2018}. Figure \ref{fig:disksize}  plots the distributions of $\tilde{R}_{\rm disk}$, where we have adopted $\varepsilon_{\rm AM}=1/3$ and $\varepsilon_{\rm disk}=0.1$. The disk size is estimated from Equation \ref{eq:rdisk} at the time when the central maximum density of the core reaches $10^{-14} \ {\rm g\ cm^{-3}}$. In Figure \ref{fig:disksize}, the average of the distribution in the case of $B_0=10 \ {\rm \mu G}$ and $B_0=2 \ {\rm \mu G}$ are 11 AU and 27 AU, respectively. The standard deviations of the distributions in the case of $B_0=10 \ {\rm \mu G}$ and $B_0=2 \ {\rm \mu G}$ are $\sigma\left(\log _{10}\left(\tilde{R}_{\mathrm{disk}}/(1 \ \mathrm{AU})\right)\right) = 0.51$ and $0.54$, respectively (Figure \ref{fig:disksize}). Note that these dispersions are independent of $\varepsilon_{\rm AM}$ because the $\varepsilon_{\rm AM}$ is constant. \par
\begin{figure}
\epsscale{0.9}
\plotone{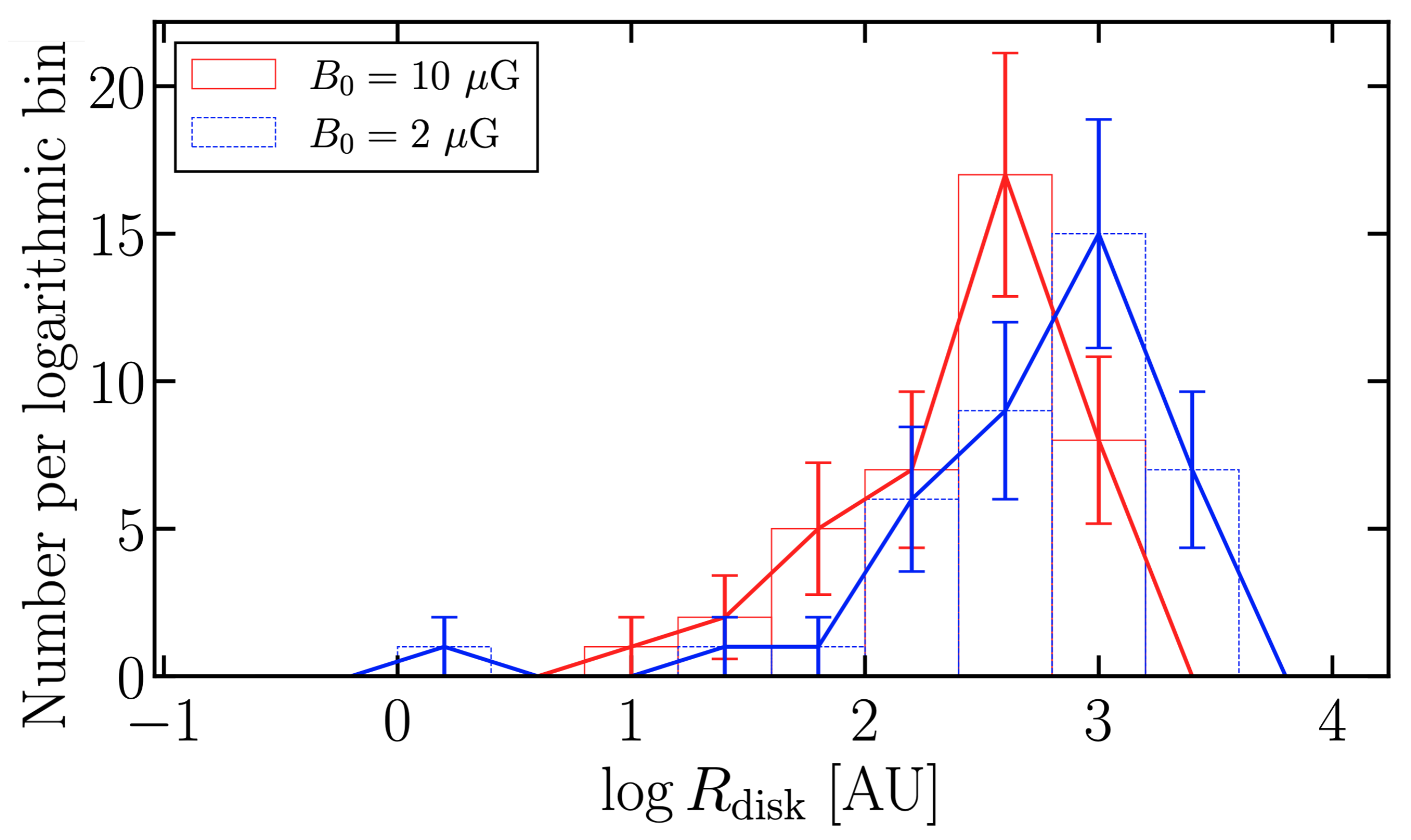}
\caption{Same as in Figure \ref{fig:disksize}, but for $R_{\rm disk}$ derived from Equation \ref{eq:rdiskshell}. \label{fig:disksizeori} }
\end{figure}

In addition, it is worth estimating the disk size formed from a part of the gas of the envelope by assuming the conservation of the angular momentum. Therefore, we also define the outer radius of the disk at its birth by the following equation: 
\begin{eqnarray}
R_{\rm disk} = \frac{j_{\rm shell}^2 }{GM_{\rm star} }.
\label{eq:rdiskshell}
\end{eqnarray}
We measure the specific angular momentum, $ j_{\rm shell}$, of the most outer concentric shell region with mass of $M_{\rm disk}$. In Figure \ref{fig:disksizeori}, the average of the distribution of $R_{\rm disk} $ in the case of $B_0=10 \ {\rm \mu G}$ and $B_0=2 \ {\rm \mu G}$ are 444 AU and 842 AU, respectively. The standard deviations of the distributions of $R_{\rm disk} $ in the case of $B_0=10 \ {\rm \mu G}$ and $B_0=2 \ {\rm \mu G}$ are $\sigma\left(\log _{10}\left(R_{\mathrm{disk}}/(1 \ \mathrm{AU})\right)\right) = 0.50$ and $0.62$, respectively. Our results indicate that the initial diversity of the angular momentum inherited from the turbulent velocity filed could explain the observed variety of the disk radii. Since we adopt the angular momentum at the final state of our simulations as $J_{\rm tot}$, the distributions in Figure \ref{fig:disksize} and \ref{fig:disksizeori} include the effect of the magnetic field on the angular momentum transfer during the core formation and gravitational collapse phases. The fact that the dispersion of the distribution does not strongly depend on the strength of the magnetic field indicates that the angular momentum transfer due to the magnetic field during the core formation and collapse stages does not have a large impact on the diversity of the disk size. Note that, since in our model $\varepsilon_{\rm AM}$ does not depend on the strength of the magnetic field, our model does not include the effect of the magnetic field on the angular momentum transfer during the disk formation phase.\par 

Note that the present estimation is based on the assumption that each core produces only one protostar. If a binary or a higher-order multiple system is created, the radius of protoplanetary disk around each star should be smaller than our estimate. Since the specific angular momentum is a slowly increasing function of the radius in the rotationally supported disk (see Equation \ref{eq:rdisk} and \ref{eq:rdiskshell}), the estimated disk size is significantly affected by the value of the specific angular momentum. In addition, as shown in Equation \ref{eq:rdisk}, the disk size is proportional to the square of $\varepsilon_{\rm AM}$. This means that variations of $\varepsilon_{\rm AM}$ will also create larger scatters in $\tilde{R}_{\rm disk}$. The results shown in \cite{Hirano2020} indicate that the resultant disk size from gravitational collapse of molecular cloud cores changes by a factor of 3 or 4 depending on the angle between the initial magnetic field line and rotation axis even if the magnitude of the angular momentum is the same at the initial state. Moreover, \cite{Tsukamoto2015b} conducted simulations of cloud core collapse with the Hall effect and shows that the Hall effect also has a large impact on the disk size. When the angular momentum vector and the initial magnetic field are in the anti-parallel configuration, the resultant circumstellar disk radius is $\sim 20$ AU at the protostar formation epoch. On the other hand, the disk with a size of $\sim 1$ AU is formed in the case of the parallel configuration. These results in  \cite{Tsukamoto2015b} suggest that the Hall effect can change the disk size by an order of magnitude at the protostar formation epoch. Although these effects may tend to increase the diversity of disk radius, the dispersion of the distribution of disk radius from the initial turbulent velocity field in the filaments alone is large enough to reproduce the observed diversity of disk size. To understand which mechanism is dominant for the diversity of disk properties, we need to solve the evolution from the filament to the circumstellar disk including non-ideal MHD effects, which is a one of our future works and beyond the scope of this paper.

\section{Summary} \label{sec:summary}
In this paper, we perform three-dimensional magnetohydrodynamic simulations using SPH to investigate the evolution of the core angular momentum in magnetized molecular filaments. The filament is perpendicularly threaded by the magnetic field with the strengths of $2 \ {\rm \mu G}$ and $10 \ {\rm \mu G}$ at the initial state. The former and latter are referred to as weak and strong magnetic filed cases, respectively. Our results are summarized as follows.\par
\begin{enumerate}
\item The internal structure of the core angular momentum depends on the strength of the initial magnetic field (Section \ref{subsec:internalAM}). We find that the dependence on the strength of the initial magnetic field can be seen only for $M \lesssim 3.0 \ {\rm M_{\odot}}$. In the outer region with enclosed mass larger than $3.0 \ {\rm M_{\odot}}$, $j$-$M$, and the $j$-$r$ relations follow the profiles inherited from the initial Kolmogorov turbulent velocity field of the filament.
\item We trace the trajectory of the SPH particles and study the evolution of the angular momentum (Section \ref{subsec:evototalAM}). The angular momentum of the cores decreases by around 30\% and 50\% during the filament fragmentation phase at the mass scale of $1\ {\rm M}_{\odot}$ for the weak and strong magnetic field cases, respectively. This result supports the conclusion of \cite{Misugi2019} in which the Kolmogorov turbulence in filaments is an origin of the angular momentum of cores.
\item We also investigate the transfer mechanism of angular momentum by analyzing each component of the torque exerted on the core (Section \ref{subsec:AMtransfer}). In the strong magnetic field case, our results show that the magnetic braking is the dominant process to remove angular momentum only for $M \lesssim 3.0 \ {\rm M_{\odot}}$. This is the reason why the angular momentum profiles for enclosed mass $M \gtrsim 3.0 \ {\rm M_{\odot}}$ are similar for both weak and strong magnetic field cases.
\item The anisotropy of angular momentum transfer is also studied in this paper (Section \ref{subsec:anisoAMtransfer}). In the strong magnetic field case,  the sample of the cores is divided into two groups: one is a perpendicular sample in which the angle between the local magnetic field and the rotation vector is larger than $60^{\circ}$, and the other is a parallel sample composed of the cores with the angle between the local magnetic field and the rotation vector smaller than $30^{\circ}$ at $  \rho_{\rm c} = 10^{-18} \ {\rm g \ cm^{-3}} $. The resultant angular momentum of perpendicular sample is smaller than that of the parallel sample by a factor of 2 at the mass scale $1\ {\rm M}_{\odot}$. This is consistent with the conclusion of \cite{Tsukamoto2018} that the perpendicular configuration transfer the angular momentum more efficiently in isothermal collapse phase compared to the parallel configuration. 
\item Since the angular momentum of the perpendicular sample is smaller than that of the parallel sample only by a factor of 2 at the final state, the distribution of the angle between the local magnetic field and the rotation vector does not show the strong alignment (Section \ref{subsec:misalignment}). In the strong magnetic field case, both the rotation and local magnetic field directions tend to be confined in the $x$-$y$ plane perpendicular to the filament long axis. This means that some of the cores in the perpendicular sample still have the perpendicular configuration just before the first core formation. Our analyses show that the strongly pinched magnetic field plays an important role to change the direction of the angular momentum from perpendicular to parallel configuration. 
\item We perform simple synthetic observations to derive the magnetic direction on the plane of the sky (Section \ref{subsec:synobs}). Our result is compatible with observations of \cite{Eswaraiah2021} which shows that the magnetic fields within the cores are randomly oriented with respect to the filament longitudinal axis. Our results suggest that this randomness is inherited from the formation of the cores from the fragmentation of magnetized filaments. 
\item It is expected that the fluid elements with larger angular momentum fall onto the central high density region when the initial magnetic field is perpendicular to the angular momentum vector as reported in \cite{Tsukamoto2018}. We find that the specific angular momentum of the perpendicular sample is the same as that of the parallel sample (Section \ref{subsec:sele}). This is because, since the degree of elongation along the $z$-axis (filament axis) is larger than that along the $x$-axis (initial direction of the magnetic field), the effect of the elongation of the cores due to the filament geometry ($z$-axis) has larger impact compared with the effect of the elongation due to the magnetic field ($x$-axis).
\item We estimate the timescale of the magnetic braking based on the concept of \cite{Mouschovias1979, Mouschovias1980}, and \cite{Mouschovias1985} (Section \ref{subsec:anatraindis}). When the model is applied to the case of the strong magnetic field, the critical mass defined as the upper limit of core mass above which the magnetic braking becomes ineffective is $\simeq 2\ {\rm M}_{\odot}$. Our analysis also shows that the filament has a critical mass  even if the Alfv\'en wave reaches infinity because the moment of inertia has an upper limit even in the perpendicular configuration. 
\item Recent observations reveal that the circumstellar disks around Class 0/\Rnum{1} has a diversity in size \cite[e.g.,][]{Tobin2020}. We estimate the distribution of disk size from the angular momentum profile around the density peak at the final state of our simulation which is just before the first core formation (Section \ref{subsec:diversity}). Our results suggest that the variety of angular momentum caused by the difference of initial phase of turbulence could be responsible for the diversity of the disk size. 
 \end{enumerate} 
As a summary, although the magnetic field plays a crucial role in the formation of filamentary molecular clouds \citep{Inoue2018,Abe2021}, the magnetic field cannot halt the subsequent fragmentation of the filament with the initial magnetic field strength of $B_0=2 \ {\rm \mu G}$ and $B_0=10 \ {\rm \mu G}$ \cite[see also][]{Hanawa2017}. In addition, the magnetic field does not play an important role at the filament scale ($\gtrsim 0.1$ pc) in terms of the angular momentum transfer. In this paper, we do not take into account the non-ideal MHD effects and the accretion onto the filament. These effects will be included in subsequent papers. 

\begin{acknowledgments}
We thank the referee for carefully reading this paper and for their constructive comments. Y. Misugi thanks Dr. Kazunari Iwasaki for providing an unpublished paper to improve our SPM code. Numerical computations were carried out on Cray XC50 at Center for Computational Astrophysics, National Astronomical Observatory of Japan. This work is supported by Grant-in-aids from the Ministry of Education, Culture, Sports, Science, and Technology (MEXT) of Japan (18H05436). This work is also supported by JSPS KAKENHI Grant Number 23K19073.
\end{acknowledgments}

%

\vspace{5mm}





\appendix

\section{Moment of Inertia in our Model}\label{sec:app1}
In this appendix, we show the equations of inertia moment used in Section \ref{subsec:anatraindis}. First, the inertia moment of the core is derived as follows:

\begin{eqnarray}
\label{eq:Icore}
\nonumber I_{x, {\rm core}} & = & I_{y, {\rm core}} \\
\nonumber  & = & \displaystyle \int \rho(r) (x^2+z^2) dV\\
\nonumber  & = & 2\pi \rho_{\rm c0}H_0^5\bar{l}_{\rm core}\left[ \displaystyle \frac{1}{2} \left\{ \ln(1+\bar{R}^2_{\rm core}) - \displaystyle \frac{\bar{R}^2_{\rm core}}{1+\bar{R}^2_{\rm core}} \right\} + \displaystyle \frac{ \bar{l}_{\rm core}^2 }{3}  \frac{\bar{R}^2_{\rm core}}{1+\bar{R}^2_{\rm core}}\right]\\
& \equiv & I_{\rm core},
\end{eqnarray}
where $\bar{R}_{\rm core}$ and $\bar{l}_{\rm core}$ represent $R_{\rm core}$ and $l_{\rm core}$ normalized by $H_0$, respectively. Note that the inertia moment of the perpendicular configuration is the same as that of the parallel configuration. The inertia moment of surrounding medium for the perpendicular configuration is described as follows:

\begin{eqnarray}
\label{eq:Iextperp}
\nonumber I_{\perp} & =  &\int \rho(r) (x^2+z^2) dV \\
\nonumber & =  &4\rho_{\rm c}H_0^5\bar{l}_{\rm core}\left[ \int^{\bar{R}_{\rm core}}_0 \left\{ \frac{\tan^{-1}\left(\bar{l}_{\rm tot}/\sqrt{1+Y^2}\right) }{\sqrt{1+Y^2}} - \frac{\bar{l}_{\rm tot}}{ 1+\bar{l}_{\rm tot}^2+Y^2 }\right\}dY  \right. \\
& + & \left. \frac{\bar{l}_{\rm core}^2}{3}\int^{\bar{R}_{\rm core}}_0 \left\{  \frac{\bar{l}_{\rm tot}}{ (1+Y^2)(1+\bar{l}_{\rm tot}^2+Y^2) } + \frac{\tan^{-1}\left(\bar{l}_{\rm tot}/\sqrt{1+Y^2}\right) }{(1+Y^2)^{3/2}}\right\}dY
\right],
\end{eqnarray}
where $\bar{l}_{\rm tot} (t) \equiv \bar{l}_{\rm A} + \bar{R}_{\rm core}$, $dX = dx/H_0$, and $dY = dy/H_0$. $\bar{l}_{\rm tot} (t)$ is derived from the following equation:
\begin{eqnarray}
\label{eq:traveltimeinapp}
\nonumber t & = & \displaystyle \int^{\bar{l}_{\rm tot}}_{\bar{R}_{\rm core}}\frac{ds}{v_{A}(s)} \\  
& = & \displaystyle \frac{H_0}{v_{\rm A}(0)} \left[ \arctan (\bar{l}_{\rm tot}) - \arctan (\bar{R}_{\rm core}) \right].
\end{eqnarray}

 The inertia moment of surrounding medium for the parallel configuration can be calculated from the following equation:
\begin{eqnarray}
\label{eq:Iextpara}
\nonumber I_{\parallel} &=& \int \rho(r) (y^2+z^2) dV \\
\nonumber & = & 4\rho_{\rm c}H_0^5\bar{l}_{\rm core}\left[ \int^{\bar{l}_{\rm tot}}_0 \left\{ \frac{\tan^{-1}\left(\bar{R}_{\rm core}/\sqrt{1+X^2}\right) }{\sqrt{1+X^2}} - \frac{\bar{R}_{\rm core}}{ 1+\bar{R}_{\rm core}^2+X^2 }\right\}dX \right. \\
&+ & \left. \frac{\bar{l}_{\rm core}^2}{3}\int^{\bar{R}_{\rm core}}_0 \left\{  \frac{\bar{l}_{\rm tot}}{ (1+Y^2)(1+\bar{l}_{\rm tot}^2+Y^2) } + \frac{\tan^{-1}\left(\bar{l}_{\rm tot}/\sqrt{1+Y^2}\right) }{(1+Y^2)^{3/2}}\right\}dY
\right].
\end{eqnarray}
By solving the equation $I_{\parallel} = I_{\rm core}$ and $I_{\perp} = I_{\rm core}$, we can derive the magnetic braking timescale.


\bibliography{paper3}{}
\bibliographystyle{aasjournal}



\end{document}